\documentclass[12pt]{article}
\pdfoutput=1
\usepackage[body={180mm,250mm}]{geometry}
\usepackage{amsfonts,amsmath,amssymb}

\newlength{\abstractwidth}
\usepackage{epsf}
\usepackage{hyperref}
\usepackage{a4wide}
\usepackage{cases}
\usepackage{epsfig}
\usepackage{color}

\newcommand{\cN}{{\mathcal N}}
\usepackage[margin=1cm]{caption}

\newcommand{\beq}{\begin{equation}}
\newcommand{\eeq}{\end{equation}}
\newcommand{\bea}{\begin{eqnarray}}
\newcommand{\eea}{\end{eqnarray}}
\def\no{\nonumber}

\newcommand{\p}{\partial}

 \newcommand{\rf}[1]{(\ref{#1})}

\begin{document}
\setcounter{footnote}{0}
\vspace{6mm}

\date{}

\title{{\Large Holographic Duals of $D=3$  $\cN=4$ Superconformal Field Theories}}

\author{ \\
{\large Benjamin~Assel,$^{\natural}$
  Costas~Bachas,$^\natural$
    John~Estes,$^\flat$ and
    Jaume~Gomis$^\dagger$ }}

\maketitle

\vskip -8.5cm
\rightline{LPTENS-11/21}
\vskip 8.5cm

\centerline{$^\natural$ Laboratoire de Physique Th\'eorique de l'\'Ecole Normale Sup\'erieure, }
\centerline{24 rue Lhomond, 75231 Paris cedex, France}

\vskip 5mm

\centerline{$^\flat$ Instituut voor Theoretische Fysica, Katholieke Universiteit Leuven, }
\centerline{Celestijnenlaan 200D B-3001 Leuven, Belgium}

\vskip 5mm

\centerline{$^\dagger$ Perimeter Institute for Theoretical Physics, }
\centerline{Waterloo, Ontario, N2L2Y5, Canada}

 \vskip 18mm

\abstract{ \normalsize{

\vskip 4mm

We find the warped $AdS_4\ltimes K$ type-IIB
 supergravity solutions holographically dual to
 a large family  of three dimensional $\cN=4$ superconformal field theories labeled by a pair $(\rho,\hat\rho)$ of partitions  of $N$.
 These superconformal   theories
arise as renormalization group fixed points  of three dimensional mirror symmetric quiver gauge theories,
denoted by $T^{\rho}_{\hat \rho}(SU(N))$ and $T_{\rho}^{\hat \rho}(SU(N))$ respectively.
We give a supergravity derivation of the conjectured field theory constraints that   must be satisfied  in order for these  gauge theories to
  flow to a non-trivial supersymmetric fixed point in the infrared.
  The exotic global symmetries of these superconformal field theories
  are precisely realized in our explicit supergravity description.  }}

\vfill\eject

\tableofcontents

\baselineskip=15pt
\setcounter{equation}{0}
\setcounter{footnote}{0}

\renewcommand{\theequation}{\arabic{section}.\arabic{equation}}

\section{Introduction}
\setcounter{equation}{0}


 The gauge/gravity duality conjecturally imprints the
  dynamics of a $D$-dimensional conformal field theory (CFT)  in the physics of string/M-theory with  asymptotically
  $AdS_{D+1}\ltimes K$ warped boundary conditions. The information about the specific CFT  is encoded in the geometry of the internal
  manifold $K$, the fluxes supporting the  $AdS_{D+1}\ltimes K$ background,
   together with the possible   presence of branes or singularities in the geometry.
  In the celebrated paper by Maldacena \cite{Maldacena:1997re}  the string/M-theory backgrounds for the
  maximally supersymmetric conformal field theories in $D=3,4$ and $6$ were identified.
   In particular, the M-theory background $AdS_{4}\times S^7$ was advanced as the holographic bulk description of the three-dimensional  $\cN=8$ superconformal field theory arising in the extreme   infrared limit of maximally supersymmetric
    three-dimensional  Yang-Mills with gauge group $SU(N)$.

 In this paper we construct the  warped $AdS_{4}\ltimes K$ backgrounds  of type-IIB string theory dual to a rich family
 of three dimensional $\cN=4$ superconformal field theories labeled by a pair $(\rho,\hat\rho)$ of partitions of $N$.
 These superconformal field theories arise
 as renormalization group fixed points   of the three dimensional $\cN=4$ mirror symmetric
 gauge theories $T^{\rho}_{\hat \rho}(SU(N))$  and  $T_{\rho}^{\hat \rho}(SU(N))$
  introduced by Gaiotto and Witten \cite{Gaiotto:2008ak} and further analyzed recently in \cite{Nishioka:2011dq}.
   These  gauge theories --
  which can be described elegantly  in terms of {\it linear} quiver diagrams -- are completely characterized by
  the choice of two partitions $\rho$ and $\hat \rho$ of $N$.

It was conjectured  in \cite{Gaiotto:2008ak}
that $T^{\rho}_{\hat \rho}(SU(N))$  and  $T_{\rho}^{\hat \rho}(SU(N))$ flow to a non-trivial
infrared  fixed point   whenever $\rho$ and $\hat \rho$   satisfy the inequality
(see section \ref{sec:quiver} for details)
  \beq
  \hat \rho^T >  \rho\,\   \  \Longleftrightarrow\  \ \rho^T> \hat \rho\,.
  \label{fixedpoint}
\eeq
When this condition is satisfied, the corresponding superconformal field theory   is invariant
 under the superconformal symmetry group $OSp(4|4)$  and  is furthermore  expected to have as global symmetry
  \beq
 H_\rho\times H_{\hat \rho}\,,
 \eeq
 where $H_\sigma$ is the commutant of $SU(2)$ in $U(N)$ for the embedding
$
 \sigma: SU(2)\rightarrow  U(N)
$
 characterized by the partition $\sigma$ of $N$  (see section \ref{sec:quiver}).
 As we will show,  our  construction of the  dual $AdS_{4}\ltimes K$ type-IIB backgrounds  gives
 a purely gravitational derivation  of the  condition \rf{fixedpoint}    necessary
for the existence of a non-trivial
superconformal field theory.
Pleasingly, our type-IIB solutions  also realize
the expected $ H_\rho\times H_{\hat \rho}$ global symmetry of these theories.
These are  non-trivial tests of the proposed holographic duality.

 The strategy behind our construction is to consider  certain limits of the type-IIB supergravity solutions \cite{DEG1,DEG2},
    devised as gravitational descriptions of supersymmetric domain walls  \cite{Gomis:2006cu} (see also \cite{Lunin:2006xr}) of four
   dimensional $\cN=4$ $SU(N)$ super-Yang-Mills. On the field theory side, these
   domain walls have been  analyzed recently in \cite{Gaiotto:2008ak,Gaiotto:2008sa}.
 The supergravity solutions have the structure of
  an  $AdS_4\times S^2\times S^2$ spacetime fibered over a Riemann surface $\Sigma$
  with disk topology. The fiber isometries realize   geometrically   the $OSp(4|4)$ superconformal symmetry of the dual theory.
   A key observation is that there exist     limits in which
    the asymptotic $AdS_{5}\times S^5$ regions of these backgrounds  decouple, and  the
 geometries go over smoothly to  $AdS_4\ltimes K$,  where $K$ is a compact manifold with specific five-brane sources (see section  \ref{sec:soln} for details).
 The limiting geometries  provide  the gravitational description of the three-dimensional $\cN=4$ superconformal field theories
  to which the quiver gauge theories $T^{\rho}_{\hat \rho}(SU(N))$ and $T_{\rho}^{\hat \rho}(SU(N))$ flow in the infrared.
  The data characterizing a given superconformal field theory is encoded in two harmonic functions $h_1, h_2$
  on the Riemann surface $\Sigma$, and in
  particular in the singularities of these functions on the boundary $\partial \Sigma$ of the Riemann surface.
  These determine completely the dual type-IIB solution.

It has been argued recently  in \cite{Bachas:2011xa} (see also \cite{Aharony:2003qf})
that various limits of the  supergravity solutions  of \cite{DEG1,DEG2}
could be important for the problem of the localization of gravity.  We will here
 see that one  interesting class of  limits are
 those  in which  the singularities on the boundary of $\Sigma$ factorize.
   In these limits  the inequality (\ref{fixedpoint})
   is  (almost)  saturated, and the dual superconformal field theory  breaks down  to (almost)  disjoint components   which are coupled
   by ``weak links" of  the quiver diagram.
    The corresponding supergravity solutions are higher-dimensional analogs
of wormholes, {\it i.e.} they have the structure of  multiple
$AdS_{4}\ltimes K$ spacetimes connected  by  ``narrow bridges" with $AdS_{5}\times S^5$ geometry.
This is similar in spirit to  the multigraviton  proposal of  references \cite{Kiritsis:2006hy,Aharony:2006hz},
 but it also differs
from this proposal  in some significant ways; in particular,  the  discussion can be kept  semiclassical.
Another interesting limit is one in which small $AdS_5\times S^5$ throats are attached to the $AdS_{4}\ltimes K$ geometry.
We will   return to these questions in a separate publication.

 \vskip 1mm

 The plan of the rest of the paper is as follows. In section \ref{sec:quiver} we introduce the relevant quiver gauge theories, describe the conditions under
 which these flow to non-trivial superconformal field theories in the infrared,  and characterize the global symmetries in  the superconformal
limit. We also briefly review the brane construction of the quiver gauge theories  and
  give a physicist's derivation,  from brane dynamics,  of  the non-trivial  condition  (\ref{fixedpoint}).
 In section \ref{sec:soln}
    we identity a consistent limit of the type-IIB supergravity solutions of \cite{DEG1,DEG2}
      in which the asymptotic $AdS_{5}\times S^5$ regions of these solutions
 go over smoothly to $AdS_4\times {\cal B}_6$,  where ${\cal B}_6$ has the topology of  a six-dimensional  ball.
   The limiting geometries
 have an  $AdS_4\ltimes K$ warped product structure,
 with multiple  five-brane
  asymptotic regions which encode the discrete data  of this class of solutions.
  In section \ref{sec:charge} we present the subtle calculation of the
  fluxes and charges of our solutions
 in terms of this discrete data. In section \ref{sec:duality} we map the supergravity data
  to the data   $(\rho,\hat\rho, N)$ characterizing the dual superconformal field theories.  We also demonstrate that our
 supergravity solutions automatically satisfy the superconformality constraint
that was conjectured  on the basis of  field theory arguments in \cite{Gaiotto:2008ak}.
We conclude in section  \ref{sec:discuss} with a brief discussion of possible extensions of our results,
and with some comments on their relevance to the problem of localization of gravity.

While this work was being completed, there appeared reference \cite{Aharony:2011yc}
which overlaps with  parts of our work. These authors  also note that asymptotic $AdS_5\times S^5$
regions can be consistently decoupled in the solutions of \cite{DEG1,DEG2}.
They  do not, however, discuss
how to construct the gravitational description of   three-dimensional superconformal theories, and the important constraint (\ref{fixedpoint}).


 \section{$T^{\rho}_{\hat \rho}(SU(N))$,  Infrared Fixed Points and Branes}
 \label{sec:quiver}
  \setcounter{equation}{0}

\begin{figure}[]
\vspace{-2cm}
\centering
\includegraphics[width=110mm]{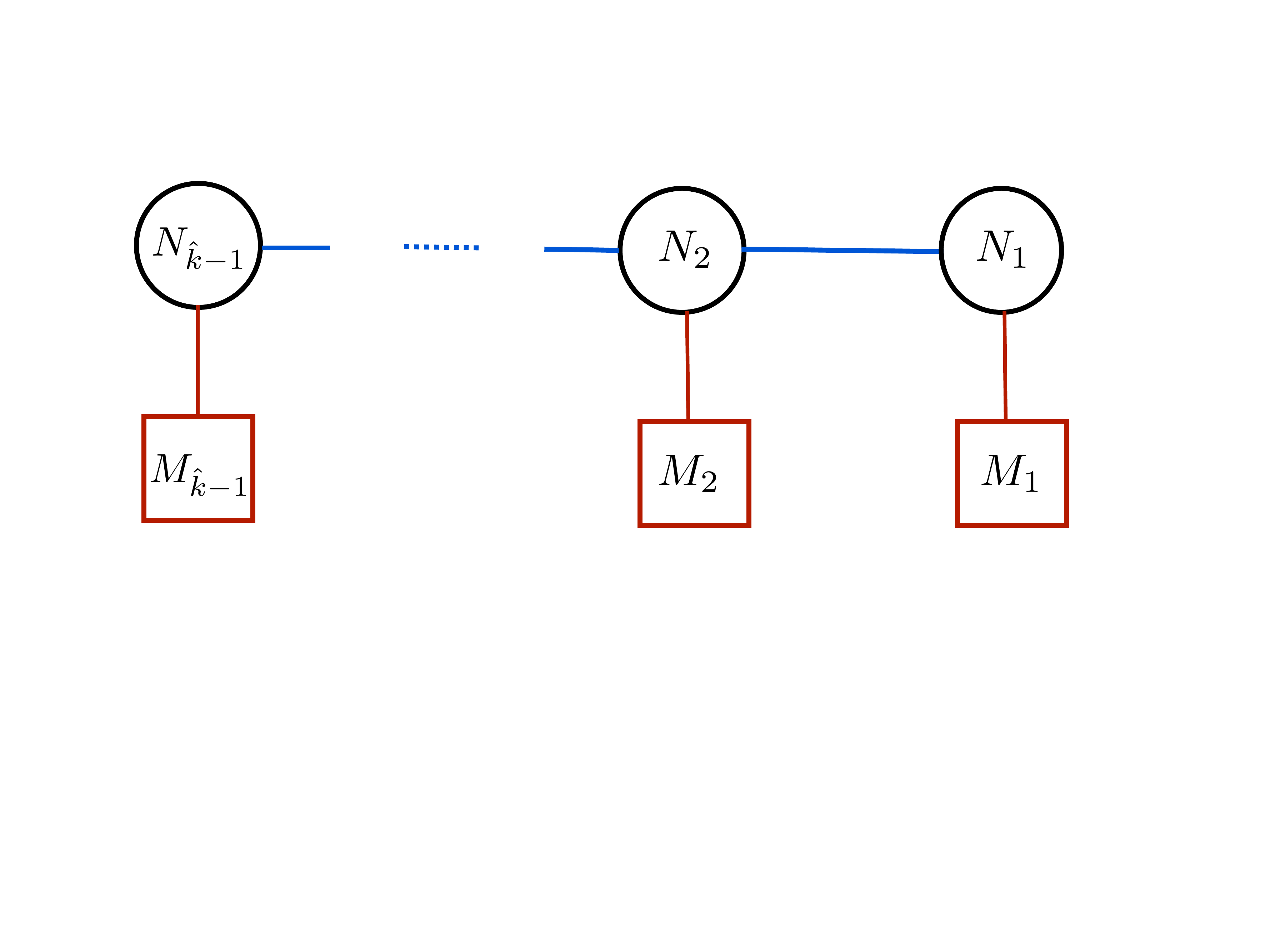}
\vspace{-3cm}
\caption{\footnotesize  A linear quiver: circles denote gauge group factors $U(N_j)$,  while squares
stand for   hypermultiplets in the  fundamental representation of the corresponding factor group.
There is also one bi-fundamental hypermultiplet for each neighbouring pair of gauge group factors, denoted by a
single blue line. }
\label{Linearquiver_General}
\end{figure}

 In  \cite{Gaiotto:2008ak}  Gaiotto and Witten
   introduced the
   theories $T^{\rho}_{\hat \rho}(SU(N))$
    which, whenever they satisfy the constraints  discussed below,
were argued to flow in the infrared to non-trivial three dimensional $\cN=4$ superconformal field theories.\footnote{For a general gauge group $\cal G$, $T^{\rho}_{\hat \rho}(\cal G)$  defines
the low energy limit of $\cN=4$ super-Yang-Mills with gauge group $\cal G$ on an interval in the presence of a duality wall and supersymmetric boundary conditions at each
end labeled by $\rho$ and $\hat\rho$, denoting two embeddings of $SU(2)$ into $\cal G$. Whether a three dimensional supeconformal field theory exists in the infrared can be inferred from the study of the moduli
space of vacua of this gauge theory. When the gauge group is $SU(N)$, these theories admit a  much simpler description in terms of conventional quiver gauge theories, to which we now turn.}
 These theories are labeled by a pair $\rho$ and $\hat \rho$ of partitions of $N$,
 which uniquely determine a three dimensional $\cN=4$ supersymmetric gauge theory in the ultraviolet limit.
 Its data is a gauge group $G$,
   and a representation $R$ of $G$ under which the hypermultiplets transform.
  More explicitly,  the gauge group of $T^{\rho}_{\hat \rho}(SU(N))$ is
   \beq\label{gaugeg}
 G=U(N_1)\times U(N_2)\times \ldots\times U(N_{\hat k-1})  \,,
 \eeq
there is one  hypermultiplet    in the bifundamental representation of each
 pair of neighboring factors $U(N_i)\times U(N_{i+1})$,   as well as  $M_i$ hypermultiplets in the fundamental representation
 of each $U(N_i)$ factor of the gauge group.
  This  supersymmetric gauge theory is summarized by the   linear quiver diagram shown in figure \ref{Linearquiver_General}.

 Any two partitions $\rho$ and $\hat \rho$  of $N$
 determine completely the gauge theory data $\{N_j,M_j\}$. Two
 useful parametrizations of the partition $\rho$  are given by
 \bea\label{partitions}
  \rho: \qquad  ~N &=&  l_1+\ldots+l_{k}  \no  \\
 &=&   \underbrace{1+\ldots+1}_{M_1}\, +\, \underbrace{2+\ldots+2}_{M_2}\, +\, \ldots+ \ldots \ .
 \eea
Here the  $l_i$ are positive non-increasing  integers,
$ l_1\geq l_2 \ldots \geq l_k > 0$, while   $M_l$
is  the  number of times the integer $l$  occurs  in this decomposition.
The $M_l$ are thus zero or positive, and  they obey the sum rule $\sum l M_l = N$.
One can associate to $\rho$
a Young tableau  whose rows have lengths $l_1, \ldots, l_k$.
The same  parametrizations   can   also be used for the partition $\hat \rho$,
\bea\label{partitionsb}
 \hat \rho: \qquad  ~N &=& \hat l_1+\ldots +\hat l_{\hat k}\no  \\
  &=&   \underbrace{1+\ldots+1}_{\hat M_1}\, +\, \underbrace{2+\ldots+2}_{\hat M_2}\, +\, \ldots+ \ldots \,.
\eea

 With this choice of ``coordinates",   $M_j$ is  precisely the number of hypermultiplets
 in the fundamental representation of the $j$th group factor, while the rank of each
 gauge group factor  is given by
 \bea
N_1 = k - \hat l_1\, , \qquad {\rm and}\qquad N_j =  N_{j-1} + m_j - \hat l_j\qquad {\rm for}\qquad j=2, \cdots \hat k - 1\ .
\label{ranks}
\eea
Here $m_l$ counts the number of terms that are  equal or bigger   than $l$
in the first line of  equation (\ref{partitions}). Thus $m_1 = k$, and
 $m_{l+1} = m_{l} -  M_l$.   As can be easily seen,  the $m_l$
are a non-increasing sequence of positive integers defining the partition  $\rho^T$,
whose Young tableau  is the transpose of  the Young tableau of the partition
 $\rho$.
\vskip 1mm

The condition   (\ref{fixedpoint})  on the two Young tableaux  is a short-hand notation
for  the following set of strict
 inequalities:
 \beq\label{ineq}
 \rho^T>\hat\rho\   \  \Longleftrightarrow\  \ \sum_{s=1}^i m_s \ >\  \sum_{s=1}^i \hat l_s\qquad \forall i = 1,\ldots ,  l_1\,.
 \eeq
Stated in words, the total number of boxes in the first $i$ rows of the Young tableau $ \rho^T$ is strictly  larger than
the same number  in the tableau $\hat\rho$. Note that the tableau $ \rho^T$ has $l_1$ rows,
while the number of rows in $\hat\rho$ is $\hat k$. Since the two tableaux have the same total number of boxes,
namely $N$, it  follows automatically   that
  $\hat\rho$  must have more rows than  $\rho^T$, {\it i.e.}  that  $\hat k > l_1$.

As  can be seen  from equation (\ref{ranks}), the condition \rf{ineq}  is  equivalent  to requiring
  that the rank $N_i$,  for
each $U(N_i)$ gauge group factor in the quiver diagram  be a positive integer.
This condition also implies that
$M_l=0$ for $l\geq \hat k$,  so that there are no hypermultiplets corresponding to
empty gauge-group factors. Thus  $\rho^T >\hat\rho$   is necessary  in order for the
linear quiver associated to the triplet $(\rho, \hat\rho, N)$ to make sense.  Note that if one of the  inequalities
 in (\ref{ineq}) is replaced by an
equality, the quiver   breaks  into two disconnected components.
What Gaiotto and Witten have conjectured  is  that {\it all}  these quiver gauge  theories flow to
non-trivial infrared fixed points  \cite{Gaiotto:2008ak}.
The existence of the dual gravity solutions,  presented in  the following section,  is  indirect
evidence in favor of  this conjecture.

 The gauge theories $T^{\rho}_{\hat \rho}(SU(N))$ have both a Coulomb   and a Higgs branch of vacua
 parametrized, respectively,  by the vector-multiplet  and hypermultiplet expectation values. Remarkably, the gauge theory
 $T_{\rho}^{\hat \rho}(SU(N))$ has precisely the same moduli space of vacua as   $T^{\rho}_{\hat \rho}(SU(N))$,
 but with the role of Coulomb and Higgs branches exchanged.
 Since $ \rho^T>\hat\rho$ implies $ \hat\rho^T>\rho$ and vice-versa \cite{Nishioka:2011dq},
 both   $T^{\rho}_{\hat \rho}(SU(N))$  and $T_{\rho}^{\hat \rho}(SU(N))$    are expected to flow to an infrared superconformal
 field theory. In fact,
   $T^{\rho}_{\hat \rho}(SU(N))$ and $T_{\rho}^{\hat \rho}(SU(N))$ are believed to flow to the same
 superconformal field theory, at the intersection of the Higgs and the Coulomb branch.
 This is a  prime example  of  three dimensional mirror symmetry \cite{Intriligator:1996ex}, which  can be attributed
 to the  S-duality of the underlying type-IIB string theory, where these theories admit an elegant brane realization (see section \ref{subsect:branecon}).

 An important guide in the construction of the dual geometries is that they must
  realize   the global symmetries of these superconformal field theories.
 The three dimensional $\cN=4$ superconformal algebra is $OSp(4|4)$. The bosonic symmetries include $SO(2,3)$,  the conformal group in three dimensions, and $SO(4) \simeq SU(2)_1\times SU(2)_2$, which is the associated $R$-symmetry.

 These superconformal field theories also exhibit, however,  a rich pattern of additional global symmetries,
 that depend on $\rho$ and $\hat \rho$ -- the data that determines as we just saw
  the (mirror pair of)  ultraviolet gauge theories whose infrared limit we want to describe.
In a given
 ultraviolet Lagrangian description, only part of the fixed-point symmetry is manifest.
 This is the one acting on the Higgs branch of the theory.  From the symmetry  acting on the Coulomb
  branch, only the  maximal abelian subgroup is in general manifest in the Lagrangian description.
  Fortunately, the Coulomb branch symmetry of a given theory -- say   $T^{\rho}_{\hat \rho}(SU(N))$ --  can be
 read from the Higgs branch symmetry of its mirror, that is   $T_{\rho}^{\hat \rho}(SU(N))$.
Thus, the full global symmetry at the superconformal   point is  believed to be $H_{\rho}\times H_{\hat \rho}$, where
   \beq
    H_{\rho}=\prod_i U(M_i)  \qquad {\rm and}\qquad  H_{\hat \rho}= \prod_i U(\hat M_i) \,.
    \eeq
This is the symmetry that rotates  the fundamental hypermultiplets of each gauge group factor  in the  quiver diagram of
$T^{\rho}_{\hat \rho}(SU(N))$ and its
 mirror.\footnote{Recall that $M_l$ was  the number of times
the integer $l$  occurs  in the partition $\rho :  N = \l_1 + \cdots + l_k$. One  may associate to this partition
an embedding of $SU(2)$ in $U(N)$,  such that the fundamental representation of $U(N)$  breaks into irreducible
components of dimension $l_i = 2s_i + 1$, where $s_i$ is the $SU(2)$  spin of the $i$th component.
It follows  that  $H_{\rho}$ is the commutant of   $SU(2)$
in $U(N)$,  for the above embedding, as has been stated in the introduction.}

 \subsection{Brane construction of linear quivers}
 \label{subsect:branecon}

The gauge  theories $T^{\rho}_{\hat \rho}(SU(N))$
can be realized  as
 low-energy  limits    of  certain   type-IIB brane configurations on the interval.
  Indeed this is how  these theories were
introduced in the first place.
We  will here sketch the salient features of these brane constructions, and refer the reader to
  \cite{Gaiotto:2008ak} and  \cite{Hanany:1996ie}   for further explanations and more references.
The  basic setup consists of: \\
\indent  \hskip 0.2cm -  a set of   $k$ D5-branes spanning  the dimensions (012456), \\
\indent  \hskip 0.2cm   - a set of  $\hat k$  NS5-branes spanning the dimensions (012789), and \\
\indent  \hskip 0.2cm  -  a set of D3-branes stretched among the five-branes along (0123).  \\
Such configurations preserve  1/4 of the  supersymmetries of type-IIB  string theory, which correspond  to the
 three dimensional
 $\cN=4$  Poincar\'e supersymmetries of $T^{\rho}_{\hat \rho}(SU(N))$. The brane configuration has
 a manifest $SU(2)_1\times SU(2)_2$ rotation symmetry in the (456) and (789)  dimensions,
 which gets identified with the $R$-symmetry   of $T^{\rho}_{\hat \rho}(SU(N))$,
 and also  coincides with the $R$-symmetry of the infrared superconformal field theory.
The vector multiplets live on the D3-branes,  and since these have finite extent on the interval along the $x^3$  dimension,  they  give rise
at large distances to a three dimensional gauge theory. Hypermultiplets
arise from open strings stretching between the D3-branes  and the D5-branes, or between two stacks of D3-branes ending on the
same NS5-brane from the left and right. The five-branes are localized in the interval direction spanned by $x^3$.

In  the infrared limit, the distance between the five-branes becomes irrelevant, and the gauge theory
is expected to flow, under suitable conditions,  to a three dimensional (in general strongly-interacting)  theory
with superconformal symmetry  $OSp(4 \vert 4)$.
The relevant  data in  the brane construction
is the ordering of the five-branes along the $x^3$ segment,
as well as the net number of D3-branes ending on each one of them.
 This data is actually redundant, since  rearrangements of
the five-branes   change  the phase  of the gauge theory, but presumably not
its superconformal (infrared) limit.    The truly relevant data,
besides the total  numbers $k$ and $\hat k$ of  D5- and NS5-branes of each kind,
are the linking numbers associated with each five-brane  separately \cite{Hanany:1996ie}.
These can be defined as follows:
\bea\label{linkingno}
l_i &=&    - n_i +  R_i^{\rm NS5}  \ \ \ \ (i= 1, \cdots k) \no\\
\hat l_j &=&   \hat  n_j + L^{\rm D5}_j \ \ \ \ (j= 1, \cdots \hat k)\,,
\eea
where
$n_i$ is the  number of D3-branes ending on the $i$th D5-brane from the right minus the
number of D3-branes ending on it from the left,  $\hat n_j$ is the same quantity for the $j$th
NS5-brane, $R_i^{\rm NS5}$ is the number of NS5-branes lying to the right of the $i$th  D5-brane,
and $L^{\rm D5}_j$ is the number of D5-branes lying  to the left of the $j$th NS5-brane.\footnote{Our definitions
differ from those in  \cite{Hanany:1996ie}   by   irrelevant signs and constant shifts.}
  The linking numbers are
 invariant under five-brane moves,  because when a D5  moves past a NS5 in the direction from left  to right,
 a D3-brane stretching between
 the two is created.\footnote{This phenomenon can be related by a chain of dualities
 to the two-dimensional anomaly equation on a D9/D1 intersection   \cite{Bachas:1997ui}. The $s$-rule,
 discussed right below,  follows   from the fact that the
 lowest-lying open string stretching between the D9- and the D1-brane is a fermion \cite{Bachas:1997kn}.}
Consistency requires that the inverse move should result in the annihilation of a stretched D3-brane
(or the creation of an anti-D3 brane).

\begin{figure}[]
\centering
\includegraphics[width=100mm]{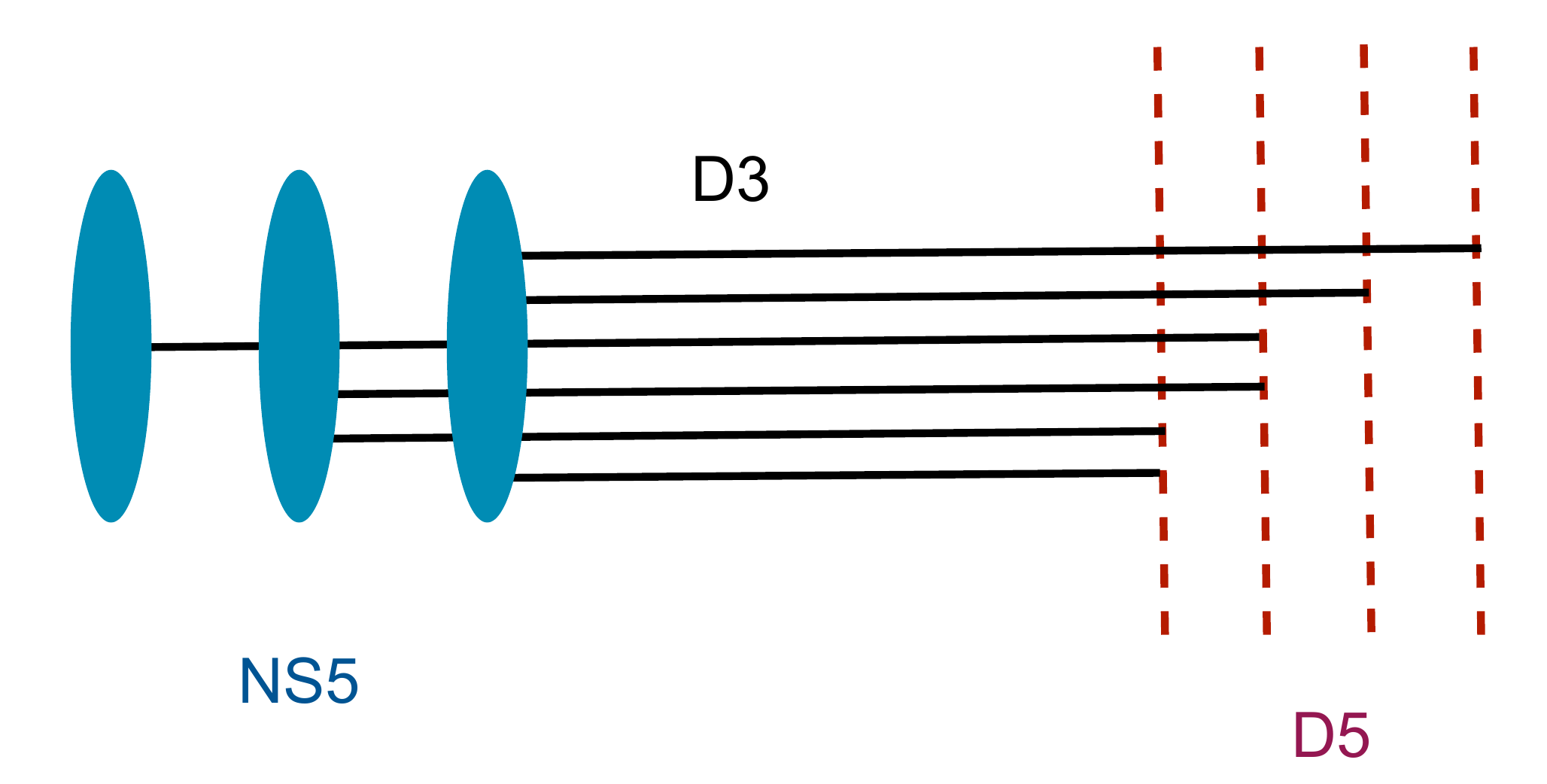}
\caption{\footnotesize   A brane configuration with $N=6$, $\rho = (2,2,1,1)$ and $\hat\rho = (3,2,1)$ . }
\label{linquiver}
\end{figure}

  The configurations of interest \cite{Gaiotto:2008ak} which realize $T^{\rho}_{\hat \rho}(SU(N))$  can     be depicted as  $N$  D3-branes in the
  middle ending on the left on a collection of NS5-branes and on the right on a collection of D5-branes.
Each D3-brane  terminates on some NS5-brane on the  left, and on some D5-brane  on the right
  (see figure \ref{linquiver}).
 This implies that the number of D3-branes that terminate on each five-brane is   precisely the linking number
  defined in \eqref{linkingno}, and that furthermore
     \bea
  N\, =\, l_1+\ldots+l_{k } \, =\,  \hat l_1+\ldots+ \hat l_{\hat k } \ .
  \eea
These are the two partitions of $N$,  $\rho$ and $\hat\rho$,   that label the   theory $T^{\rho}_{\hat \rho}(SU(N))$.
The partition $\hat\rho$ encodes the linking numbers of the NS5-branes,  while $\rho$ encodes  the linking number of the D5-branes.
$S$-duality of the type-IIB string theory exchanges the two type of five-branes, and thus  the two partitions. Therefore, the S-dual brane configuration
realizes the mirror theory $T_{\rho}^{\hat \rho}(SU(N))$, so that S-duality acts as mirror symmetry in the gauge theory.

Since five-branes of the same kind are not connected with D3-branes, they  can be moved freely past each other,
so their  relative order
is irrelevant in the infrared limit. We will adopt the convention that the linking numbers of the NS5-branes are
non-decreasing from left to right, and that the same holds for the D5-branes from right to left. Thus $l_1\geq l_2
\cdots \geq l_k$ and $\hat l_1\geq \hat l_2 \cdots \geq \hat l_{\hat k}$, where $i=1$ and $ j=1$ are the innermost
five-branes, while $i=k$ and  $j= \hat k$ are the outermost ones. As was already explained, it is convenient to   associate to $\rho$ a
Young tableau whose rows have $l_1, \cdots , l_k\, $ boxes, and likewise for   $\hat \rho$.
\vskip 1mm

\begin{figure}[]
\vspace{1cm}
\begin{center}
\begin{tabular}{c @{\hspace{0.5in}} c}
\includegraphics[height=1.5in]{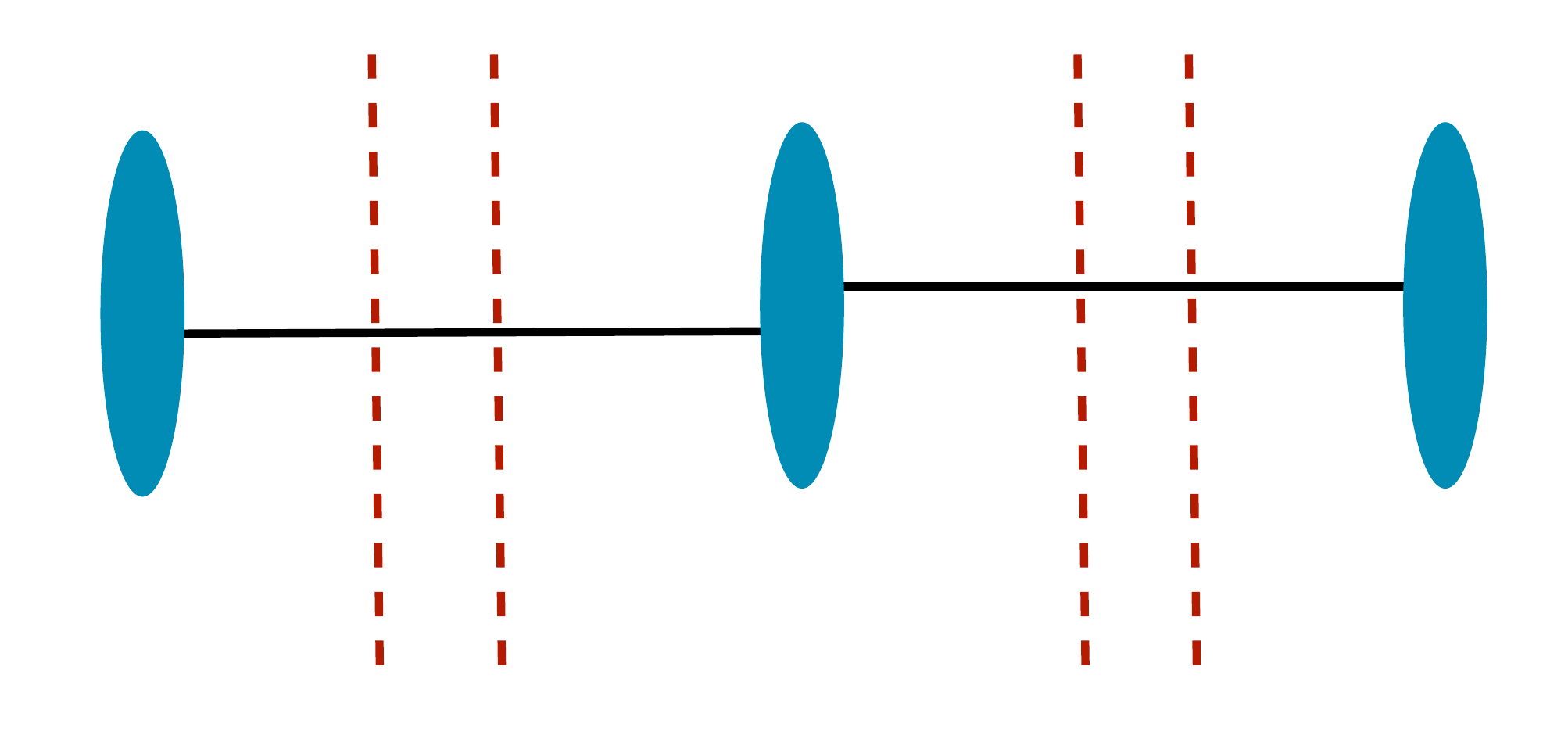} &
\includegraphics[height=1.3in]{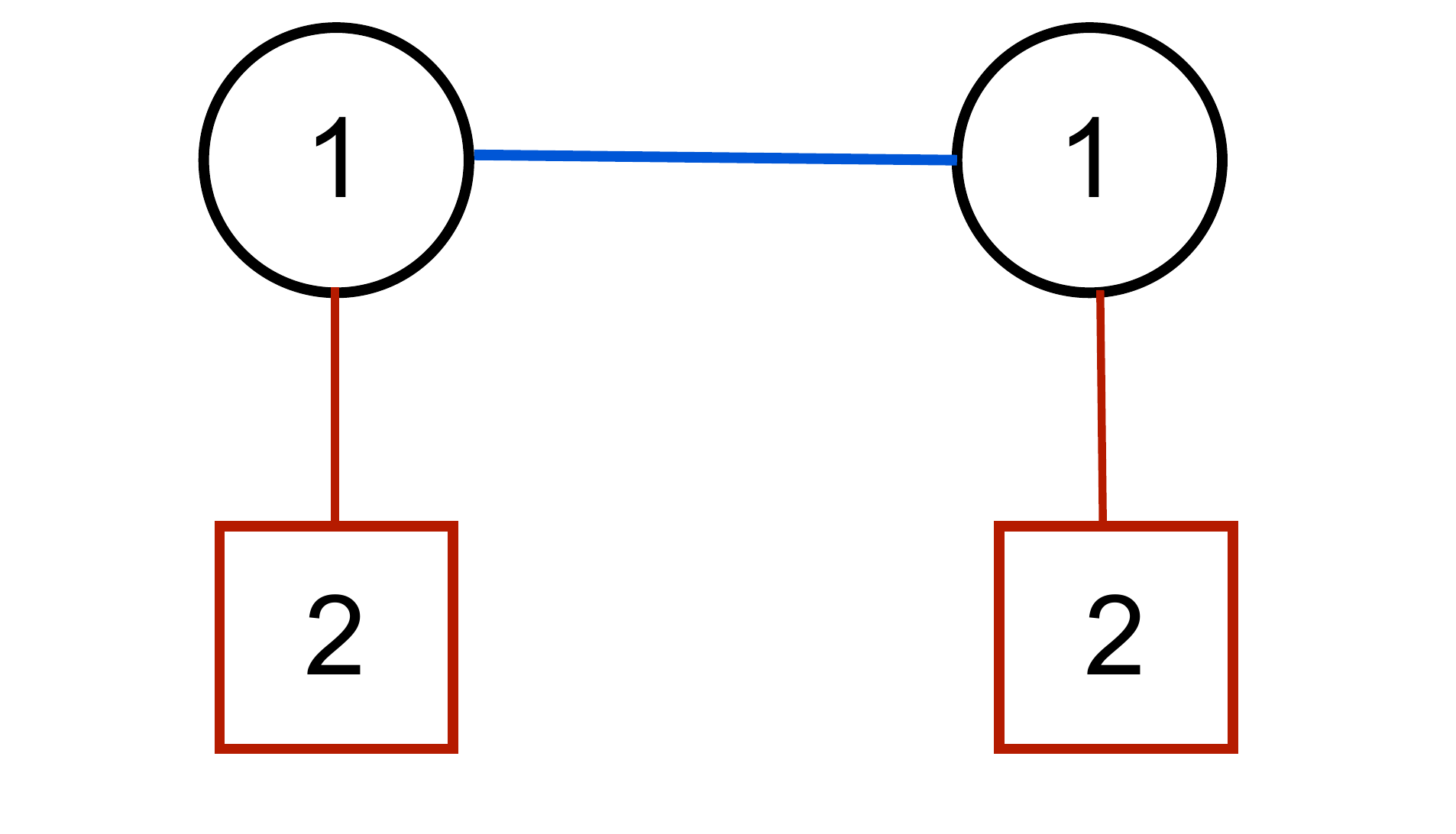}
\end{tabular}
\end{center}
\caption{\footnotesize  On the left, the brane construction  of figure 2 after moving the D5-branes in the way described
in the text. On the right the quiver diagram describing  the corresponding supersymmetric gauge theory.}
\label{Linearquiver2}
\end{figure}

 We are now ready to explain the   condition  that forces the ordering  (\ref{ineq})
of the Young tableaux. To understand the origin of this constraint,  let us  try to rearrange all   D5-branes
  so that no D3-branes terminate on them at all, i.e. so that their linking numbers are equal to the number of
  NS5-branes that lie on their right. This is the configuration in which the field content of the quiver gauge theory
  is most easy to read (see figure \ref{Linearquiver2}). The argument that we will now explain shows, in a nutshell,
  that unbroken supersymmetry
  requires the weaker condition $\hat \rho^T \geq  \rho$, or equivalently $\rho^T \geq  \hat \rho$. However, when the inequality is saturated the
  corresponding quiver gauge
  theory breaks down to  pieces that flow to non-interacting superconformal theories in the infrared.

 To see why,  start moving   the innermost
D5-brane  towards  the left. Each time it crosses a NS5, one of the D3-branes ending on it is  destroyed.
If $l_1 < \hat k$ the process  stops  when  there are no more D3-branes terminating on  our D5-brane,
{\it i.e.} after having crossed all but $\hat k - l_1$ of the  NS5-branes.  What if $l_1 >  \hat k$? In this case,
  after moving $\hat k -1$ steps,   our D5-brane will be attached to the outermost NS5-brane by more than one
 stretched D3-branes.
 This is forbidden  by the  {\em s-rule},  which states that such a gauge theory
 would have no supersymmetric ground states \cite{Hanany:1996ie}.   The marginal case $l_1 = \hat k$ is not forbidden
 by the $s$-rule. In this case, however,  in the final step the D5-brane will be detached
 from the rest of the quiver, and  the  infrared superconformal theory could  be described by a
 partition  $\rho^\prime$  whose Young tableau has one less row than our original partition $\rho$.  Since we only want to characterize
 distinct superconformal theories, we should not count separately $\rho$ and $\rho^\prime$.  By convention we only
 keep the partitions with the minimal number of rows.  This implies the additional condition of $\hat\rho^T > \rho $ namely $\hat l_1 < k$, which ensures that no NS5-branes are detached to the right of the quiver.

 Assume then  that $l_1 < \hat k$, and try now to rearrange the second innermost D5.
 The first two D5-branes have a total of $l_1 + l_2$ D3s ending on them.
 At most $\hat M_1 = \hat m_1 - \hat m_2$ of these can terminate on NS5-branes that have only a single D3-brane attached.
[Recall the definition of $\hat m_l$ as the number of NS5-branes with at least $l$ D3-branes attached, or equivalently
as the partition corresponding to the transposed Young tableau $\hat\rho^T$].
  There remain, therefore,  \emph{at least}   $l_1 + l_2 - (\hat m_1 - \hat m_2)$   D3-branes, that must be attached to the remaining
 NS5-branes. Unless
 \bea\label{2ndineq}
2 \hat m_2 \geq   l_1 + l_2 - (\hat m_1 - \hat m_2) \ \Longleftrightarrow \  \hat m_1 +  \hat m_2 \geq l_1 + l_2\ ,
 \eea
 some  D5/NS5 pairs would be attached by more than one D3-brane, thereby violating the $s$-rule.
 Thus supersymmetry requires that  (\ref{2ndineq}) hold.
 Furthermore,  In the marginal case $ \hat m_1 +  \hat m_2 = l_1 + l_2$,
   the quiver gauge  theory  can be again reduced, {\it i.e.} it
   breaks down to separate   pieces that flow to decoupled superconformal theories in the infrared. This can be shown by first moving
   both D5-branes $\hat m_2$ steps, so that they enter the region of singly-attached NS5-branes.
   The latter are furthermore all attached to these two D5-branes, and to nothing else.  It is then easy to see that the   remaining
   $\hat m_1 - \hat m_2$ moves   will
   necessarily break up the quiver.

 \vskip 1mm

 We  now state the general result, which can be proved  by induction with the above logic.
By slight abuse of notation, we write $\rho^T \geq  \hat\rho$ for the weaker  form of
(\ref{ineq}) in which the total number of boxes up to the $i$th row must be greater {\it or  equal}  on the two sides.
Then the brane rearrangement argument shows that:
 \bea
  {\rm supersymmetry} \,  \Longleftrightarrow \,  \hat \rho^T  \geq  \rho \ ,  \ \ \ {\rm and}  \ \
  \ {\rm irreducibility}\,  \Longleftrightarrow \,  \hat \rho^T  >  \rho\ .
 \eea
   When the strong form of the constraint is satisfied, the configuration after completing all rearrangements of D5-branes
      consists of a connected linear chain of $\hat k$ NS5-branes, attached in  pairs by $N_j$ D3-branes. The D5-branes
     intersect, but are detached from the D3-branes.  The corresponding gauge theory data can be read easily
     from this configuration: there is one $U(N_j)$ gauge group factor for every set of stretched D3-branes,  one hypermultiplet
     in the bifundamental representation for each adjacent pair, and one hypermultiplet in the fundamental representation
     of the corresonding gauge group
     for each D5-brane. The final result agrees  precisely with  the  gauge theory content of $T^{\rho}_{\hat \rho}(SU(N))$
     described in the previous subsection.


 \section{The ${\rm AdS}_4\ltimes K$  Supergravity Solutions}
  \label{sec:soln}
  \setcounter{equation}{0}

We  will now  derive
 the gravitational backgrounds dual to the superconformal field theories labeled by $(\rho,\hat \rho)$
described in the previous section,   as special limits
of the   type IIB-supergravity solutions found in  \cite{DEG1,DEG2}.  These were analyzed as candidate
backgrounds for gravity localization in \cite{Bachas:2011xa}, whose   conventions
and notation we adopt.  The main new result in this section is the existence of a smooth
limit in which the asymptotic ${\rm AdS}_5\times {\rm S}^5$ regions of the solutions of \cite{DEG1,DEG2} are truncated away, and
the space transverse to the ${\rm AdS}_4$ slices is compactified to $K$.

\subsection{Local solutions: General form}

 For the reader's convenience we collect here the formulae describing the general form of the solutions of \cite{DEG1,DEG2}.
  These solutions   are fibrations
of  ${\rm AdS}_4\times {\rm S}^2\times  {\rm S}^2$  over a base space which is a Riemann surface $\Sigma$ with the
topology of a disk.  The general discussion of these solutions \cite{DEG2}  is most convenient with a choice of complex coordinate
 that varies over the upper-half plane, but for our purposes here we prefer to use a coordinate that varies over
 the infinite strip:  $$\Sigma \equiv  \{ z \in \mathbb{C}\, \vert \, 0\leq {\rm Im}z \leq {\pi\over 2}\}\ . $$
 The solutions have a manifest   ${SO}(2,3) \times {SU}(2)_1 \times{SU}(2)_2$ symmetry realized on the fibers, which combined  with the  sixteen super(conformal) symmetries of the solutions, give a bulk realization of the three dimensional $OSp(4|4)$ supeconformal algebra.
 The solutions are  completely specified by two  functions $h_1(z, \bar z)$ and $h_2(z, \bar z)$ which are real
 harmonic and regular inside $\Sigma$, and which obey the boundary conditions (here $\partial_\perp$
 is the normal derivative):
\bea\label{bconditions}
h_1 = \partial_\perp h_2 = 0 \hskip 0.5cm {\rm for} \hskip 0.5cm  {\rm Im}z=0\ ,  \hskip 1cm
h_2 = \partial_\perp h_1= 0 \hskip 0.5cm {\rm for} \hskip 0.5cm  {\rm Im}z= {\pi\over 2}\ .
\eea

In writing down the solutions one also needs the dual harmonic functions,
 which are defined  by the following  relations
\bea
\label{dualharmfunc}
h_1 = -i({\cal A}_1 - \bar {\cal A}_1) \qquad\rightarrow  \qquad  h_1^D  = {\cal A}_1 + \bar {\cal A}_1\ ,  \no\\
h_2 = {\cal A}_2 + \bar {\cal A}_2 \qquad \rightarrow \qquad  h_2^D = i({\cal A}_2 - \bar {\cal A}_2)\ .
\eea
The dual functions are in general ambiguous. These ambiguities  will
have, however,  a simple physical interpretation,  as  gauge transformations of the RR and NSNS 2-form gauge potentials
(see section \ref{sec:charge}). Besides the dual functions, it is also
  convenient to define the following combinations of $h_1$, $h_2$,  and of their first derivatives
(here $\p = \p/\p z, \bar\p = \p/\p\bar z$) :
 \bea\label{W}
W &=& \p  h_1 \bar\p  h_2 + \bar\p h_1 \p  h_2  = \p \bar\p  (h_1h_2)\ ,  \no\\
N_1 &=& 2 h_1 h_2 |\p  h_1|^2 - h_1^2 W \ , \no \\
N_2 &=& 2 h_1 h_2 |\p  h_2|^2 - h_2^2 W   \ .
\eea
Now in  the conventions of  \cite{Bachas:2011xa} the solution reads:
\vskip 0.1mm
\bea\label{metric}
\underline{\rm Metric}:
\qquad ds^2 = f_4^2 ds_{{\rm AdS}_4}^2 + f_1^2 ds_{{\rm S}_1^2}^2 + f_2^2 ds_{{\rm S}_2^{2}}^2 + 4 \rho^2 dz d\bar z \ ,
\eea
where
\bea\label{metricfactors}
f_4^8 &=& 16\, {N_1 N_2 \over W^2}\ ,
\qquad \qquad
\rho^8 = {N_1 N_2 W^2 \over h_1^4 h_2^4}\ ,  \no\\
f_1^8 &=&  16\,  h_1^8 {N_2 W^2 \over N_1^3}\ ,
\qquad  \qquad
f_2^8 = 16\,  h_2^8 {N_1 W^2 \over N_2^3}\ ,
\eea
and the ${\rm AdS}_4$ and
2-sphere metrics are normalized to unit radius;

\bea \label{dilaton}  \hskip -6cm
\underline{\rm Dilaton}:
\qquad  e^{4 \phi} = {N_2 \over N_1}\ ;
\eea
\bea\label{3forms}
\underline{\rm 3{\rm -}forms}:  \qquad   H_{(3)}  =   \omega^{\, 45}\wedge db_1  \qquad {\rm and}\qquad
 \,  F_{(3)} =   \omega^{\, 67}\wedge db_2   \ ,
\eea
where  $H_{(3)}$ and $F_{(3)}$ are the    NS/NS  and R/R\,   3-form field strengths,
$ \omega^{\, 45}$ and $ \omega^{\, 67}$ are the volume forms of the unit-radius  spheres  ${\rm S}_1^{2}$ and ${\rm S}_2^{2}$,
and \hskip 0.5mm
 \bea\label{3forms1}
b_1 &=& 2 i h_1 {h_1 h_2 (\p  h_1\bar  \p  h_2 -\bar \p  h_1 \p  h_2) \over N_1} + 2  h_2^D \ ,  \no\\
b_2 &=& 2 i h_2 {h_1 h_2 (\p  h_1 \bar\p  h_2 - \bar\p  h_1 \p  h_2) \over N_2} - 2  h_1^D \ ;
\eea
 \bea \label{5form}
\hskip -0.6cm
\underline{\rm 5{\rm -}form}:  \qquad   F_{(5)}  =
 - 4\,  f_4^{4}\,  \omega^{\, 0123} \wedge {\cal F} + 4\, f_1^{2}f_2^{2} \,  \omega^{\, 45}\wedge \omega^{\, 67}
 \wedge (*_2  {\cal F})   \ ,
\eea
where $ \omega^{\, 0123}$ is the volume form of the unit-radius ${\rm AdS}_4$,
${\cal F}$ is a 1-form on $\Sigma$ with the property  that  $f_4^{\, 4} {\cal F}$ is closed,
and $*_2 $ denotes Poincar\' e duality with respect to the $\Sigma$ metric.
 The explicit expression
for ${\cal F}$  is given by
\bea
f_4^{\, 4} {\cal F} = d j_1\   \qquad {\rm with} \qquad j_1 =
3 {\cal C} + 3 \bar  {\cal C}  - 3 {\cal D}+ i \frac{h_1 h_2}{W}\,   (\p  h_1 \bar\p  h_2 -\bar \p h_1 \p h_2) \ ,
\eea
where ${\cal C}$ is defined by the relation $\p {\cal C} = {\cal A}_1 \p  {\cal A}_2 - {\cal A}_2 \p  {\cal A}_1$
while ${\cal D} = \bar {\cal A}_1 {\cal A}_2 + {\cal A}_1 \bar {\cal A}_2$.\footnote{Note
 that the corresponding expressions (9.61) and (9.63) in \cite{DEG1} are missing the factor of ${\cal D}$.}

\vskip 1mm

 The above set of expressions  gives the local form of the general  solution
 for  the ansatz of  references  \cite{DEG1,DEG2}.
These  expressions are invariant under  conformal transformations of the coordinate $z$,
which  map,   however,  in general $\Sigma$ to a different disk-like domain of the complex plane.


\subsection{Asymptotic $AdS_5 \times S^5$ regions and five-branes}

The simplest solution with all necessary ingredients for our purposes here corresponds to the following choice of
real harmonic functions:    \cite{DEG2,Bachas:2011xa}
\bea\label{hNS5D5}
h_1 &=&    \, \left[ -i \alpha \, {\rm sinh} (z - \beta ) -    \gamma\,  {\rm ln}\left(  {\rm tanh} \left({i\pi\over 4} - {z - \delta \over 2}\right)\right)
\right]
 +  {\rm c.c.} \ ,  \no\\
h_2 &=&  \, \left[ \hat \alpha  \,  {\rm cosh}(z- \hat\beta )   -   \hat \gamma \,
 {\rm ln}\left({\rm tanh} \left({z - \hat\delta \over 2}\right)\right)\right]  +  {\rm c.c.}   \  .
\eea
The parameters ($\alpha,\beta,\gamma,\delta$)  and ($\hat\alpha,\hat\beta,\hat\gamma,\hat\delta$)
are all real. The only other condition on this set of parameters,
explained in  \cite{DEG2},  is that  $\alpha\gamma$ and $\hat\alpha\hat\gamma$  must  be non-negative.
If not,  the solution has curvature singularities supported  on a one-dimensional curve in the interior of $\Sigma$,
which have no interpretation in string theory.

This solution  describes the near-horizon geometry of stacks of
 intersecting D3-branes,  NS5-branes  and D5-branes. The setup preserves the same
 super-Poincar\'e and $R$ symmetries  as the brane constructions   considered
  in our discussion of linear quiver gauge theories
 in section \ref{sec:quiver}.
Flipping the sign of $\alpha$ and $\gamma$ amounts to trading the D3-branes and D5-branes for anti-branes.
Without loss of generality, we will thus assume from now on
 that $\alpha, \gamma, \hat\alpha , \hat\gamma$ are all non-negative.

Let us describe the main features of the above background.
In the two regions $Re(z) \rightarrow \pm \infty$ it approaches asymptotically
 the $AdS_5 \times S^5$ solution,  with the values of the radius and the dilaton  given by: \cite{Bachas:2011xa}
\bea\label{radii}
 L_\pm^4\,  =\,
 16 \vert \alpha^\pm \hat \alpha^\pm \vert \, {\rm cosh} (\beta^\pm - \hat\beta^\pm)
 \qquad {\rm and}\qquad
e^{2\phi_\pm} = \left\vert {\hat\alpha^\pm \over \alpha^\pm} \right\vert e^{\pm (\beta^\pm - \hat\beta^\pm)}\ ,
 \eea
 where $\alpha^\pm$ and $\beta^\pm$  are given by:
  \bea\label{phys2}
 \alpha^\pm = \alpha\, \sqrt{1 + {4\gamma\over \alpha}e^{\pm(\delta -  \beta) }}\ ,
 \qquad  e^{\beta^\pm}  = e^{\beta}\, \left(1 + {4\gamma\over \alpha}e^{\pm(\delta - \beta) }\right)^{\pm1/2} \  ,
 \eea
with similar expressions holding for the hatted quantities $\hat\alpha^\pm$ and $\hat\beta^\pm$.
In the limit   $\gamma = \hat \gamma = 0$, the solution becomes the supersymmetric Janus domain wall,
in which the dilaton field varies  from one asymptotic $AdS_5 \times S^5$ region to the other.
Setting in addition   $\beta =\hat\beta= 0$,   gives the
 global $AdS_5 \times S^5$  solution with radius $L^4 = 16\, \alpha \hat \alpha$
 and a constant dilaton given by $e^{2\phi} =  \hat \alpha/\alpha $.

The other important   feature of the above solution
 is the presence of singularities on the boundary of the strip,
 namely at $z= i \frac{\pi}{2} +\delta$ and $z = \hat \delta$. These are   associated with the non-trivial 3-cycles in the geometry.
The solutions that we consider have the property that on the lower boundary of $\Sigma$ the two sphere $S_{1}^{2}$ shrinks to zero size smoothly (there is no conical singularity) and on the upper boundary of $\Sigma$ the two sphere $S_{2}^{2}$ shrinks to zero size smoothly.
 Therefore, with the exception of the singular points $z= i \frac{\pi}{2} +\delta$ and $z =\hat  \delta$, the rest of the boundary of the strip
  corresponds to  interior points of the ten dimensional geometry.   Now consider a small open curve $I$ in $\Sigma$, which surrounds
  the singularity at $z =\hat  \delta$ and ends on the ${\rm Im}(z)$=0 axis, where the 2-sphere $S_1^2$ shrinks to zero.
  Then  $I \times S^2_1$ is a non-contractible 3-cycle which
  is, furthermore,   threaded by non-vanishing $H_{(3)}$ flux, as can be checked with the help of
  the expressions  (\ref{3forms}) and (\ref{3forms1}). This flux signals that the local geometry
  describes a stack of  NS5-branes, whose total charge  is
  proportional to $\hat\gamma$. Likewise,  the region near $z= i \frac{\pi}{2} +\delta$ describes a stack of D5-branes,
  with total charge proportional to the parameter $\gamma$ (see section \ref{sec:charge}).
  Note that at these singularities the dilaton field diverges,  as  expected near the location of five-branes.

\begin{figure}
\center
\includegraphics[width=12cm]{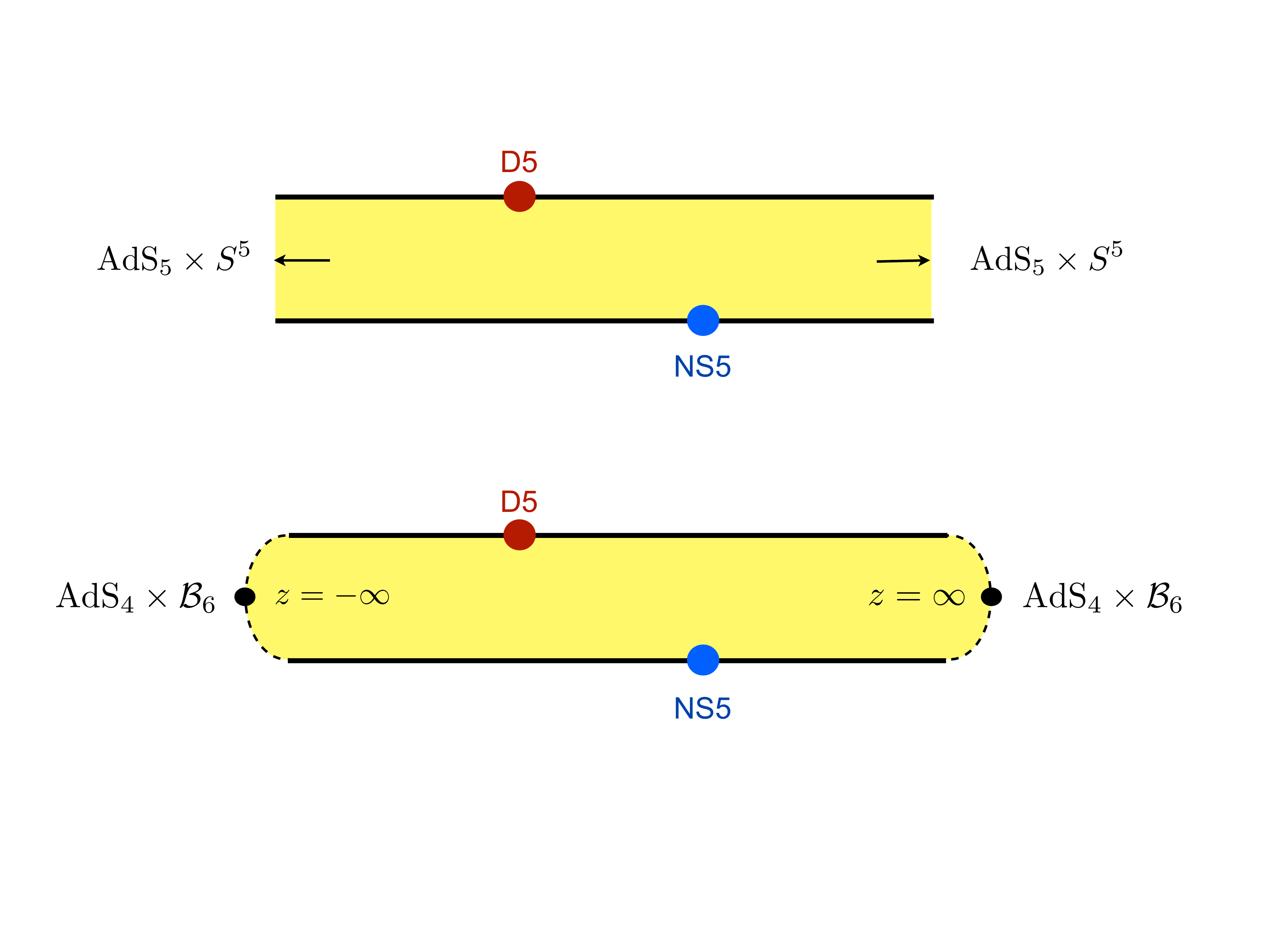}
\vskip -1.2cm
\caption{\footnotesize
The  four asymptotic regions of the solution (\ref{hNS5D5}): near  $z\simeq  \pm\infty$
the geometry asymptotes to $AdS_5 \times S^5$,  while the singularities on the lower and the upper strip boundary
describe stacks of NS5 branes and  D5 branes. Taking $\alpha$ and $\hat\alpha$ to zero replaces the
  $AdS_5 \times S^5$ regions by smooth caps  homeomorphic  to  $AdS_4$ times a 6-dimensional ball.
  }
\label{strip_2stacks}
\end{figure}

\subsection{Closing the $AdS_5 \times S^5$ regions}

 As explained in the last subsection, the $z\to\pm\infty$ regions of the strip describe regions of the 10-dimensional
 solution that approach  $AdS_5 \times S^5$,   with the radii given by the expressions (\ref{radii}) and (\ref{phys2}).
 These radii vanish
 if we take $\alpha$ and $\hat\alpha$ to zero, while keeping the other parameters of the solution  fixed.
 Interestingly enough the limit  is smooth:   the asymptotic $AdS_5 \times S^5$ regions are replaced in this  limit
 by regions that are homeomorphic to $AdS_4\times {\cal B}_6$,  where ${\cal B}_6$ is the 6-dimensional ball.
 This is depicted schematically in the lower part of figure \ref{strip_2stacks}.

Let us be a little more explicit. The limiting geometry is described by the two harmonic functions:
\bea\label{hNS5D5limit}
h_1 =  \,   -    \gamma\,  {\rm ln}\,  {\rm tanh} \left({i\pi\over 4} - {z - \delta \over 2}\right)
 +  {\rm c.c.} \ ;   \ \
h_2 =   \,   -   \hat \gamma \,
 {\rm ln}\, {\rm tanh} \left({z - \hat\delta \over 2}\right)  +  {\rm c.c.}   \  .
\eea
Inserting these two functions in the expression (\ref{metric})  for
 the metric, and making the following change of coordinates:
 \bea
  r^2 = 2 (e^{2\delta} + e^{2\hat\delta})\, e^{-2x}\qquad {\rm where} \qquad z = x+iy\ , \no
 \eea
gives in the limit    $x\to\infty$:
\bea
ds^2  \simeq
L^2\,
 \Big[ ds^{2}_{AdS_4} \, + \, dr^2 \, + \, r^2 \left( \sin^2 y \, ds^{2}_{S^2_1} \,
+ \, \cos^2 y \, ds^{2}_{S^2_2} \, + \, dy^2\right) \Big]\,
\eea
with
\bea
L^4 =   {16 \, {\gamma \hat \gamma } \over {\rm cosh}(\delta -\hat\delta) }\ .
\eea
This is  locally $AdS_{4} \times \mathbb{R}^{6}$,
which shows that the $z\simeq \infty$ region becomes a regular interior
region of the 10-dimensional geometry, as advertized.
The Ricci scalar in the $x\to\infty$ limit  asymptotes to
${\cal R} \simeq  8/L^2$, while the dilaton and the $p$-form fields are also finite.
The region $x\to -\infty$ can be analyzed similarly; it is in fact sufficient to flip the signs
of $\delta$ and $\hat\delta$ in the above expressions.

The complete metric defined by the harmonic functions  (\ref{hNS5D5limit})   describes a
warped product  $AdS_4 \ltimes K$, where $K$  is a compact 6-dimensional manifold
with admissible singularities at the location of the five-branes.\footnote{The NS5-brane geometry
is  easier to recognize
in the string-frame metric. Expanding
 near $z = \hat\delta$  one gets:  $ds_{\rm string}^{\, 2} \simeq
 4\hat\gamma
  [ du^2 + u(ds^{2}_{AdS_4} + ds^{2}_{S^2_2}) + d\theta^2 + \sin^2 \theta  ds^{2}_{S^2_1}  ] $,
 where $x-\hat\delta = y\, {\rm tan}\theta$,  and $(x-\hat\delta)^2 + y^2 = 4 e^{-u}$.
 This is the expected metric for a   NS5-brane  whose worldvolume wraps $AdS_4\times S_2^2$.
 The geometry near the D5-brane is described by the same Einstein-frame metric,
  but  opposite value of the dilaton field.
}
 Notice that the overall
scale of the metric is proportional to $\sqrt{\gamma\hat\gamma}$,  so  that both types
of five-brane stacks are required  for a regular solution. This is our first example of a
background which is the gravity dual of the $\cN=4$  superconformal theories labeled by the pair of partitions $(\rho,\hat \rho)$.
As will become clear in the following sections, this first example corresponds to two  equipartitions of the D3-branes,
 $N = lM_l =  \hat l \hat M_{\hat l}$.   The simplest possible partitions
 \beq
\rho=\hat \rho:  \qquad N= \underbrace{1+1+\ldots +1}_N\,
\eeq
are obtained in the special case
 $\gamma = \hat\gamma$ and
$\hat\delta -\delta = \ln\tan{\pi\over 2N}$.
 The superconformal theory (and quiver gauge theories) corresponding  to this simplest choice of partitions
   is sometimes denoted by just $T(SU(N))$.
Notice that for this example the  gauge theory is identical to its mirror.


\subsection{Many stacks of five-branes}

It is easy to generalize the above solution so as to include  many singularities which will describe
the asymptotic regions  of
different stacks of D5-branes and NS5-branes.  The corresponding harmonic functions are given by
\bea\label{hmany}
h_1 &=& \left[ -i\alpha  \sinh(z-\beta ) - \sum_{a=1}^{q} \gamma_{a} \ln\left(\tanh\left(\frac{i\pi}{4}-\frac{z-\delta_a}{2}\right)\right) \right] + c.c. \nonumber\\
h_2 &=& \left[ \hat \alpha  \cosh(z-\hat\beta ) - \sum_{b=1}^{\hat{q}} \hat{\gamma}_{b} \ln\left(\tanh\left(\frac{ z-\hat{\delta}_{b} }{2}\right)\right) \right] + c.c.
\eea
with $\delta_1 < \delta_2 < ... < \delta_q $ and $\hat{\delta}_1 > \hat{\delta}_2 > ... > \hat{\delta}_{\hat{q}}$.

The solution  described by these harmonic functions
  contains two asymptotic $AdS_5 \times S^5$ regions, $q$  singularities on the upper boundary of $\Sigma$
 corresponding to $q$ stacks of D5-branes,  and $\hat{q}$  singularities on the lower boundary of $\Sigma$
 corresponding to $\hat{q}$ stacks of NS5-branes.
 The $a^{th}$ stack of D5-branes is located at $z = i \frac{\pi}{2} + \delta_a$ and contains a number of D5-branes proportional to $\gamma_{a}$,
 while the $b^{th}$ stack of NS5-branes is located at $z = \hat{\delta}_b$
 and contains a number of NS5-branes proportional to $\hat{\gamma}_{b}$.
Note that we choose to label the singularities on the upper boundary of the strip from left to right, and on the
lower boundary from right to left.  This choice will prove convenient  when  identifying the parameters
of the solution
with the data of the dual superconformal field theory,   in   section \ref{sec:duality}.


\begin{figure}
\center
\vskip -1cm
\includegraphics[width=12cm]{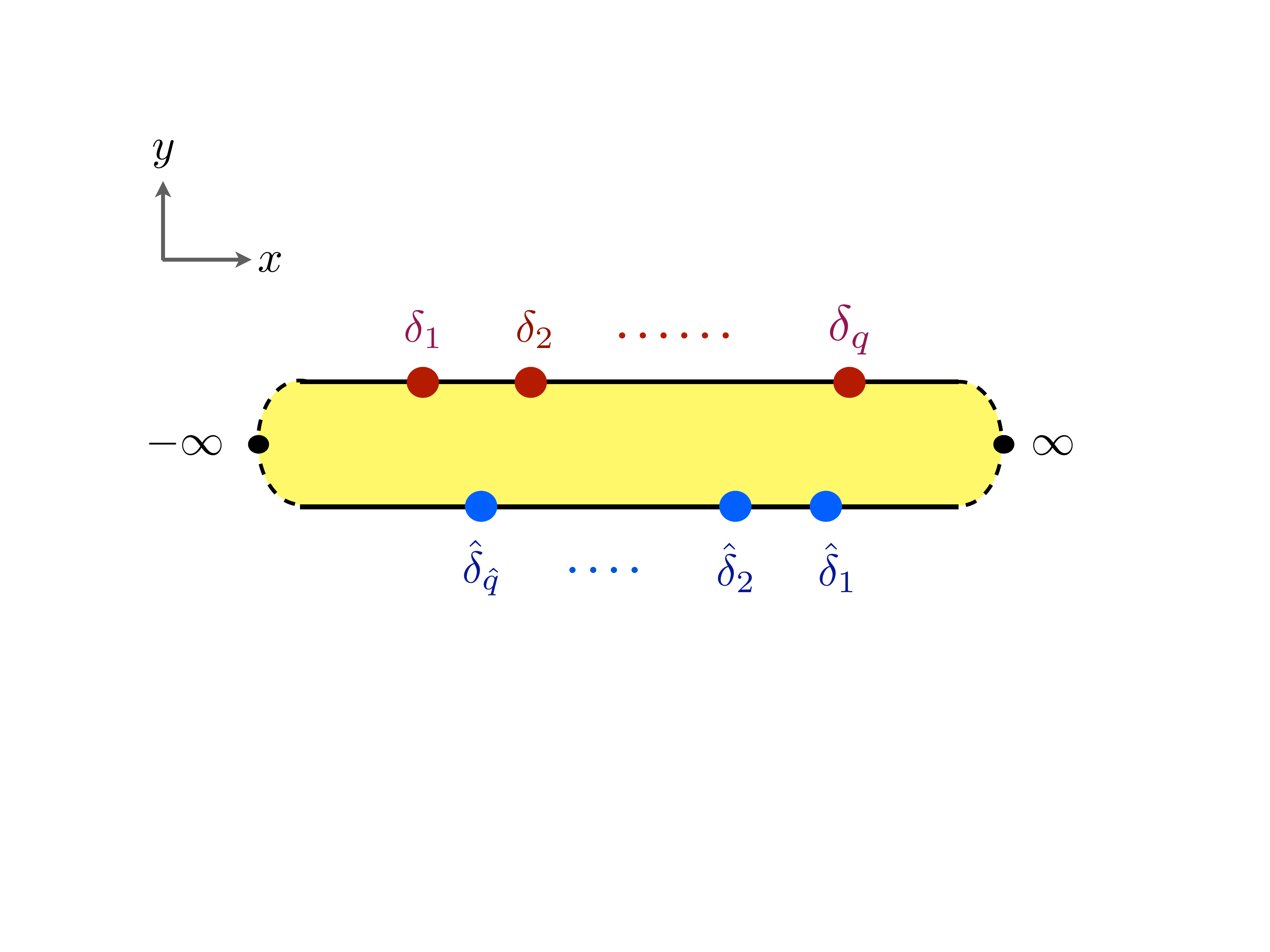}
\vskip -2.1cm
\caption{\footnotesize
The infinite strip with several singularities, corresponding to many different stacks of five-branes.
The positions of these singularities along the real axis  are related to the data of the dual superconformal
field theory,  in a way that will be detailed in section \ref{sec:charge}.
 }
\label{strip_manystacks}
\end{figure}

The solution (\ref{hmany})  describes the near-horizon geometry
 of a brane construction that contains
 D3-branes stretched between the two asymptotic regions,
 between the asymptotic regions and the stacks of five-branes,  and between the  D5- and  NS5-brane stacks.
 This can be seen from the calculation of the 5-form flux that enters in the various asymptotic
  regions, as will be detailed in the following section.

We may now proceed as before to  close the two asymptotic $AdS_5 \times S^5$ regions by setting $\alpha  = \hat\alpha  = 0 $.
The resulting harmonic functions are :
\bea
h_1 &=&    - \sum_{i=a}^{q} \gamma_{a} \ln\left(\tanh\left(\frac{i\pi}{4}-\frac{z-\delta_a}{2}\right)\right)   + c.c. \nonumber\\
h_2 &=&  - \sum_{b=1}^{\hat{q}} \hat{\gamma}_{b} \ln\left(\tanh\left(\frac{ z-\hat{\delta}_{b} }{2}\right)\right)  + c.c.
\eea
In this class of solutions the manifold is of the type $AdS_4 \ltimes K$ where the $AdS_4$ is fibered over the compact
 six-dimensional manifold $K$. The closure of the $AdS_5 \times S^5$ regions is smooth,  and the points at infinity become
  interior points (locally $AdS_4 \times \mathbb{R}^6$), as explained in the previous subsection.
In the rest  of this paper we will  focus on this class of type-IIB supergravity solutions,
and propose a precise correspondence with the three dimensional $\cN=4$ superconformal field theories
 labeled by the two partitions $(\rho,\hat \rho)$, and  discussed in section \ref{sec:quiver}.

We should   mention here as a side remark
 that it is also  possible to close only one asymptotic $AdS_5 \times S^5$ region,
  by taking an appropriate limit of parameters.
   In this case it seems natural, even if we didn't look at it in detail, that one can derive a similar correspondence with four dimensional
$\cN=4$ super-Yang-Mills  on  half-space with suitable half-supersymmetric  conditions imposed at the boundary.
 This type of boundary conditions has been studied in detail  in  \cite{Gaiotto:2008sa,Gaiotto:2008ak}.
 The supergravity analysis of this case has appeared  in the recent reference  \cite{Aharony:2011yc},
 which  has some partial overlap with our  work. The possibility of closing off  an $AdS_5 \times S^5$ region
 has been, in particular, also observed in this reference. Here we close off both $AdS_5 \times S^5$ regions, and
  end up with $AdS_4\ltimes K$ backgrounds dual to three dimensional superconformal field theories,
  rather than  $\cN=4$ super-Yang-Mills in half-space. Another interesting class of  limits are factorization limits
  of  the five-brane singularities; these
  will be discussed   briefly near the end of this paper.


\section{Brane Charges and Quantization}
\label{sec:charge}

In order to discuss the precise correspondence between the
supergravity solutions of section \ref{sec:soln} and the superconformal field theories to which $T^{\rho}_{\hat \rho}(SU(N))$  and
$T_{\rho}^{\hat \rho}(SU(N))$ flow in the infrared,
 we must first compute the charges   contained in the supergravity backgrounds.
The definition of the amount of D3-brane charge dissolved into the D5-brane and NS5-brane stacks
is subtle and suffers from a well known ambiguity.  In particular, the so-called  ``Page charge"
(which is  quantized and localized)
 transforms under large gauge transformations of the two-form gauge potentials $B_2$ and $C_2$
 (see \cite{Marolf:2000cb} for a nice discussion of the issue).\footnote{The fully  gauge-invariant D3-brane charge is   a non-linear and non-local
 functional of the supergravity fields,   reflecting the (still  partially-understood)
 non-abelian nature of the underlying gauge symmetries.
This  charge has been  computed by  exact worldsheet techniques for the NS5/D3-brane system in
 \cite{BDS}. }
  We will
 give below a physical interpretation of this gauge ambiguity  in terms of the Hanany-Witten effect \cite{Hanany:1996ie}.

  Let us start then by introducing the non-trivial 3- and 5-cycles which support the D3-brane, D5-brane and NS5-brane charges:
\begin{itemize}
\item ${\cal C}^a_3$: are defined by the fibration of $S_2^2$ over a line segment in $\Sigma$
 which ends on the upper boundary of the strip and encloses the point $\delta_a$.
  Note that $h_2 = 0 \Rightarrow f_2 = 0$ on the upper boundary, so that  ${\cal C}^a_3$ is topologically also a 3-sphere.
\item $\hat {\cal C}^b_3$:
are defined by the fibration of $S_1^2$ over a line segment in $\Sigma$
 which ends on the lower boundary and encloses the point $\hat \delta_b$.  Note that
 $h_1 = 0 \Rightarrow f_1 = 0$ on the lower boundary so that $\hat {\cal C}^b_3$ is topologically a 3-sphere.
\item ${\cal C}^a_5$:
 is defined by the warped product $S_1^2 \ltimes {\cal C}^a_3$ and is topologically an $S^3 \times S^2$.
\item $\hat {\cal C}^b_5$: is defined by the warped product $S_2^2 \ltimes \hat {\cal C}^b_3$ and is topologically an $S^3 \times S^2$.
\end{itemize}
Recall that $a=1,\ldots,q$  and $b=1,\ldots,\hat q$,
where $q$ and $\hat q$ are  the number of D5-  and NS5-brane stacks
 in the supergravity solution that is determined by the two harmonic functions \rf{hmany}.
 The orientation of the cycles is chosen in such a way  that  the line segments on $\Sigma$ are always oriented
 counter-clockwise.

In evaluating the brane charges, we shall need the expressions for the two dual harmonic functions:
\begin{align}
\label{dualfs}
h_1^D &= \zeta + \left[ \alpha  \sinh(z-\beta) - i \sum_{a=1}^{q} \gamma_{a} \ln\left(\tanh\left(\frac{i\pi}{4}-\frac{z-\delta_a}{2}\right)\right) \right]
 + c.c. \ ; \cr
h_2^D &= \hat \zeta +
\left[ i \hat\alpha \cosh(z-\hat\beta) - i \sum_{b=1}^{\hat{q}} \hat{\gamma}_{b} \ln\left(\tanh\left(\frac{ z-\hat{\delta}_{b} }{2}\right)\right) \right] + c.c. \ .
\end{align}
These expressions are ambiguous, because the imaginary part of the logarithmic
function $f= \gamma {\rm log} z$ depends
on the choice of  the branch cut (whereas its real part  is unambiguous
everywhere other than  at $z=0$).
 In general, a different choice can be made for each $a$ and $b$, but the most natural choice
is to put  {\it all}  logarithmic  cuts outside the infinite strip $\Sigma$.
With this choice, $h_1^D$ has a discontinuity of  $2\pi \gamma_{a}$ at the $a$-th singularity
on the upper boundary of the strip, and $h_2^D$ has a discontinuity of
$- 2\pi\hat  \gamma_{b}$ at the $b$-th singularity
on the lower boundary.
This leaves a residual ambiguity,
which has to do with the
choice of the phases at infinity; it is  parameterized  by the real constants $\zeta$ and $\hat \zeta$ in the above expressions.
  The meaning of these ambiguities in the choice of $h_1^D$ and $h_2^D$  will become clear shortly.

The five-brane charges are defined in the standard way:
\begin{align}
\label{5branecharges}
Q^{{(a)}}_{D5} &= \int_{{\cal C}^a_3} F_3 = (4 \pi)^2 \gamma_a = 4 \pi^2 \alpha^\prime N^{(a)}_{D5} \cr
\hat Q^{{(b)}}_{NS5} &= \int_{\hat {\cal C}^b_3} H_3 = -(4 \pi)^2 \hat \gamma_b = -4 \pi^2 \alpha^\prime \hat N^{(b)}_{NS5} \ ,
\end{align}
where $N^{(a)}_{D5}$ is the number of D5-branes in the $a$-th D5-brane stack and $\hat N^{(b)}_{NS5}$ is the number of
NS5-branes in the $b$-th NS5-brane stack. These charges are local, gauge invariant and conserved.
We have used also here the fact that they are
 quantized in units of $2 \kappa_0^2 T_5$,
  where $2 \kappa_0^2 = (2 \pi)^7 (\alpha^\prime)^4$ is the gravitational coupling constant, and
  $T_5 = 1/[(2 \pi)^5 (\alpha^\prime)^{3}]$ is the five-brane
   tension. Note that since we have kept the dilaton arbitrary,  we are free to  set the string coupling $g_s=1$;
   the tension of the NS5-branes and the D5-branes is thus the same.

\smallskip

Because of the presence of five-branes,
 the definition of the D3-brane charge is more subtle.
A   ``brane-source charge" can be defined by the failure of the Bianchi identity for the gauge-invariant field strength
\cite{Marolf:2000cb}
\begin{align}
dF_5 - H_3 \wedge F_3 &= \ast j^{\rm bs}_{D3} \ .
\end{align}
However, in the presence of either D5- or NS5-branes, $j^{\rm bs}_{D3}$ is not conserved, since
\begin{align}
d (\ast j^{\rm bs}_{D3}) = - (\ast j_{NS5}) \wedge F_3 +  H_3\wedge (\ast j_{D5})   \,,
\end{align}
where $d H_3 = \ast j_{NS5}$ and $d F_3 = \ast j_{D5}$.
As a result  the brane-source charge is neither localized nor   conserved.
 It is possible, however,  to introduce a conserved current, which we shall denote by $j^{\rm Page}_{D3}$,
 at the cost of gauge invariance:
\begin{align}
\ast j^{\rm Page}_{D3} = \ast j^{\rm bs}_{D3}\,  + \,    (\ast j^{}_{NS5})\wedge C_2  -  B_2 \wedge (\ast j^{}_{D5})  \ .
\end{align}
The corresponding charge
  is local, conserved and turns out to be quantized \cite{Marolf:2000cb},   but it  is not gauge invariant.
It  is usually called the Page charge.

This  rather formal argument  boils down basically to the following fact:  the Page charge is given by
the integral of either $ F^\prime_5 = F_5 + C_2\wedge H_3$  or of
$F^{\prime\prime}_5 = F_5 - B_2\wedge F_3$, both of which obey the
non-anomalous
Bianchi identity in the absence
of brane sources. Which of these two choices is the appropriate one, depends on which of the two potentials,
$B_2$ or $C_2$,  can be defined globally on the 5-cycle over which one wishes to integrate.
Consider for example  $\hat {\cal C}^b_5$ : as  can be easily verified, the integral of  $F_3$  on any
3-subcycle of $\hat {\cal C}^b_5$ is zero, so that $C_2$ can be   defined on this 5-cycle globally.
The corresponding D3-brane Page charge   therefore reads
\begin{align}
\label{D3charge1}
\hat Q^{{\rm Page} (b)}_{D3} &= \int_{\hat {\cal C}_5^b} F_5 + C_2 \wedge H_3 \ .
\end{align}
Similarly, on  the
the ${\cal C}^a_5$ 5-cycles, the gauge potential   $B_2$ can be defined globally, and
  we may thus write the D3-brane Page charge as follows:
\begin{align}
\label{D3charge2}
Q^{{\rm Page} (a)}_{D3} &= \int_{{\cal C}^a_5} F_5 - B_2 \wedge F_3 \ .
\end{align}
To make the notation lighter, we will from now on drop the word ``Page"  when we refer to a D3-brane charge. All
D3-brane charges will be  Page charges.
  \smallskip

It turns out that the only non-vanishing contribution to the D3-brane charges  comes from the Chern-Simons term,
 and we find
\begin{align}
Q^{ (a)}_{D3} =  - 4 \pi b_1\Bigl\vert_{z=\delta_a+i \pi/2} \ Q^{(a)}_{D5} \ , \qquad \
\hat Q^{ (b)}_{D3} =  4 \pi b_2\Bigl\vert_{z=\hat \delta_b} \ \hat Q^{(b)}_{NS5} \ .
\end{align}
One can understand these formulae
by taking  the integration cycles
 to lie very close to the 5-brane singularities.
 The  gauge potentials are in this case constant,
 while the integrals over the 3-form fluxes give  exactly the 5-brane charges (\ref{5branecharges}).
In terms of the parameters
appearing in the harmonic functions (\ref{hmany}), these D3-brane charges can be written explicitly as follows:
\footnote{We have made use of the identity $\arctan x = -\frac{i}{2}[\ln(1+ix)) - \ln(1-ix)]$ to simplify the formula.}
\begin{align}
\label{gencharge}
Q^{ (a)}_{D3}  &=  2^8 \pi^3\,  \left(
 \hat \alpha\,   \gamma_a \sinh(\delta_a - \hat\beta) - 2\,  \gamma_a \sum_{b=1}^{\hat q} \hat \gamma_b  \arctan (e^{\hat \delta_b - \delta_a})
  \right)  \ ,   \cr
\hat Q^{ (b)}_{D3} &=  2^8 \pi^3\,  \left(
 \alpha \,  \hat \gamma_b \sinh(\hat \delta_b - \beta) + 2\,
   \hat \gamma_b \sum_{a=1}^q \gamma_a \arctan (e^{\hat \delta_b - \delta_a})
 \right) \ .
\end{align}
The arctangent functions are   here taken to be  real.
These expressions were obtained with the  choice of logarithmic branch cuts described after equation (\ref{dualfs}),
and with   $\zeta = \hat\zeta = 0$. We will refer to this as the ``canonical gauge"   choice.
\smallskip

     Let us discuss   the choice $\zeta =0$. From equations (\ref{3forms}) it follows that
     $C_2 \sim b_2 \omega^{67}$ approaches $-2\zeta \omega^{67}$  in the $z\to -\infty$ region.
     Since the 2-sphere $S_2^2$   shrinks to zero size everywhere on the upper strip boundary, a
     gauge that is non-singular at $z\to -\infty$  must correspond to  the choice $\zeta =0$.
      With this choice  the  2-form gauge potential, $C_2$, is
      well-defined everywhere,  except on the part of  the upper boundary  of the strip
     starting from  ${\rm Re} z\geq   \delta_{1}$.  Likewise setting
    $\hat\zeta = 0$ ensures that  the  2-form gauge potential $B_2$ can be
    well-defined everywhere  in the strip, except on the lower boundary for
   ${\rm Re} z\leq \hat \delta_1$.  Thus, in the  canonical gauge for the gauge potentials,  the number of patches
   required  to cover the entire spacetime is minimal.  Other choices of the constants $\zeta$ and $\bar\zeta$,
   or different choices of the logarithmic branch cuts, would have lead to a description requiring
  more  coordinate patches.
\smallskip

We focus now on the solutions with $\alpha = \hat\alpha =0$, for which  the asymptotic  $AdS_5 \times S^5$ regions are capped off.
 Denoting the net number of D3-branes ending on the $a$-th
 D5-brane stack by $N^{(a)}_{D3}$  and  the net number of D3-branes ending on the
 $b$-th NS5-brane stack  by $\hat N^{(b)}_{D3}$,
  and using the quantization conditions for the charges, we find the following two  relations
\begin{align}
\label{genchargequant}
N^{(a)}_{D3} &=   - N^{(a)}_{D5} \sum_{b=1}^{\hat q} \hat N^{(b)}_{NS5}\,  \frac{2}{\pi} \arctan (e^{\hat \delta_b - \delta_a}) \ ,  \cr
\hat N^{(b)}_{D3} &=   \hat N^{(b)}_{NS5} \sum_{a=1}^q N^{(a)}_{D5}\,  \frac{2}{\pi} \arctan (e^{\hat \delta_b - \delta_a}) \,.
\end{align}
These formulae place restrictions on the values $\delta_a$ and $\hat \delta_b$ may take in the full quantum theory:
they must be chosen so that,  for given $N^{(a)}_{D5}$ and $\hat N^{(b)}_{NS5}$, the above formulae produce integer
numbers of D3-branes.  It is interesting to note that, taken together, the equations
 (\ref{5branecharges}) and (\ref{genchargequant}) are sufficient to quantize all the parameters
  in the supergravity solution.

Let us illustrate this point in the simplest case of  a single stack  of D5-branes  and a single stack
of NS5-branes. Dropping the indices one finds
 $\hat N_{D3} = -N_{D3}$,  and
 \begin{align}
\label{singlestacks}
\hat \delta - \delta = \ln \tan \left( \frac{\pi}{2} \frac{\hat N_{D3}}{N_{D5} \hat N_{NS5}} \right) \ .
\end{align}
The quantized parameter $\hat \delta - \delta$ becomes quasi-continuous when
$N_{D5} \hat N_{NS5}\gg 1$, {\it i.e.} for very large numbers of five-branes. The parameter
$\hat \delta + \delta$, on the other hand, is irrelevant because a real translation of the origin
of the $z$ axes does not change the supergravity solution. Counting also $\gamma$ and $\hat\gamma$,
we thus have three physical parameters quantized so as to give three integer charges.

 Expressing the parameters $\delta_a$ and $\hat \delta_b$ in terms of the integer  quantities $N^{(a)}_{D3}$,
 $\hat N^{(b)}_{D3}$, $N^{(a)}_{D5}$ and $\hat N^{(b)}_{NS5}$  in the general case is much more difficult.
Let us however do a simple counting: there are $q+\hat q$ parameters $\delta_a$ and $\hat \delta_b$,
but one of them  is irrelevant  and can be eliminated by an overall shift.
This matches the number of integer D3-brane charges in five-brane stacks, which are subject to the overall charge conservation condition
\bea\label{chargeconserv}
- \sum_{a=1}^q N^{(a)}_{D3} = \sum_{b=1}^{\hat q} \hat N^{(b)}_{D3}\
=  \sum_{a=1}^q   \sum_{b=1}^{\hat q}  N^{(a)}_{D5}\, \hat N^{(b)}_{NS5} \,  \frac{2}{\pi} \arctan (e^{\hat \delta_b - \delta_a})
\eea
This condition follows from (\ref{genchargequant})  by summing over the indices $a$ and $b$.

 In fact, since
  the arctangent functions are bounded from above by $\pi/2$, the allowed distribution of D3-brane charges is also subject to
 the following  two  inequalities:
\begin{align}
\vert  N^{(a)}_{D3} \vert \,  \leq\,  N^{(a)}_{D5} \sum_{b=1}^{\hat q} \hat N^{(b)}_{NS5}
\qquad {\rm and} \qquad
\vert \hat N^{(b)}_{D3} \vert \,  \leq\,  \hat N^{(b)}_{NS5} \sum_{a=1}^q N^{(a)}_{D5} \ .
\end{align}
  These conditions can be attributed to the $s$-rule. Note indeed that the total number of D3-branes
  emanating from the $a$-th  D5-brane stack
   cannot exceed the number of D5-branes in the stack, times the total
  number of NS5-branes. If it did exceed, some D5/NS5 pairs would be connected by more than one D3-brane,
  which would constitute a violation of the $s$-rule \cite{Hanany:1996ie}.

\smallskip

   Under large gauge transformations which change $\zeta$ and $\hat\zeta$ from zero to some finite values,
   the  integer D3-brane charges (\ref{genchargequant}) transform as follows:
\bea
\delta N^{(a)}_{D3} = - {2 \hat\zeta \over \pi\alpha^\prime}\, N^{(a)}_{D5}\ , \qquad {\rm and}\qquad  \delta \hat N^{(b)}_{D3} =
-{2 \zeta \over \pi\alpha^\prime}\,  \hat N^{(b)}_{NS5}\ .
\eea
Thus,  it is natural to define appropriate ratios  which we will refer to by anticipation as  ``linking numbers":\footnote{The
 signs are chosen  so as to agree with our earlier  convention.}
\bea\label{linking}
l^{(a)} \, \equiv\, - { N^{(a)}_{D3} \over N^{(a)}_{D5}} \, \qquad {\rm and}\qquad
\hat l^{(b)} \, \equiv\, { \hat N^{(b)}_{D3} \over  \hat N^{(b)}_{NS5}}\ .
\eea
These  transform under the large gauge transformations by constant shifts. It is actually possible to
define gauge-invariant but non-local D3-brane charges, by subtracting a contribution at infinity:
\begin{align}\label{gaugeinv}
 Q^{{\rm inv} (a)}_{D3} &= \int_{{\cal C}^a_5} F_5 - B_2 \wedge F_3 +   \int_{{\cal C}^a_3}   F_3 \int_{S_1^2}  B_2\Bigl\vert_{z=\infty}   \cr
  \hat Q^{{\rm inv} (b)}_{D3} &= \int_{\hat {\cal C}^b_5} F_5 + C_2 \wedge H_3  -  \int_{\hat {\cal C}^b_3}
    H_3  \int_{ S_2^2} C_2\Bigl\vert_{z=-\infty} \ .
 \end{align}
 It is now easy to check that  large gauge transformations,  such as a different choice for a logarithmic branch cut,
 changes the term at infinity in precisely the way needed to cancel the variation of the local charge. This is the supergravity analog
 of the Hanany-Witten effect, which trades a number of D3-branes ending on a given D5- or NS5-brane,
 for the equivalent  number of five-branes of the opposite type that have crossed to the right,
 or to the left \cite{Hanany:1996ie}.

  In the canonical gauge,  the contributions  at infinity in  definitions (\ref{gaugeinv}) vanish. Thus the linking numbers
  that we computed above
  can be considered as the gauge-invariant linking numbers.


\section{The Holographic Duality Map}
\label{sec:duality}
  \setcounter{equation}{0}

The goal of this section is to establish an explicit correspondence between the three dimensional $\cN=4$ superconformal field theories introduced in section \ref{sec:quiver}
and the $AdS_4\ltimes K$ supergravity solutions presented in section \ref{sec:soln}.

We recall that this family of superconformal field theories is labeled by the triplet   $(\rho, \hat \rho, N)$ and describe the infrared limit of the $T^{\rho}_{\hat \rho}(SU(N))$  and $T_{\rho}^{\hat \rho}(SU(N))$ quiver gauge theories.
  The partitions of $N$  labeled $\rho$ and $\hat \rho$   were identified with the
  linking numbers $l_i$ and $\hat l_j$ of D5-branes and NS5-branes  appearing in the brane construction in section \ref{subsect:branecon}.
  They obey $N = \sum_{i=1}^k l_i = \sum_{j=1}^{\hat k} \hat l_j$.  In this  brane construction, the D3-branes end on a collection of D5-branes and NS5-branes localized on an interval, and yield at low energies three dimensional gauge field theories.

  On the other hand, in section \ref{sec:soln} we have constructed type-IIB supergravity solutions with the  $OSp(4|4)$ symmetry which is necessary to yield a gravitational description of three dimensional $\cN=4$ superconformal field theories. In order for the solutions to describe three dimensional field theories on the boundary, however, we must decouple the asymptotically $AdS_5 \times S^5$ regions present in these geometries.
  Otherwise these supergravity solutions describe a four dimensional field theory
   in the presence of a boundary or domain wall. Fortunately, we have shown that a simple limit, obtained by setting
   $\alpha = \hat \alpha = 0$,  caps off these asymptotic regions and yields   a solution of the type $AdS_4\ltimes K$, precisely as required for three dimensional conformal field theories.

  In section  \ref{sec:charge}, we have defined the supergravity analog of the linking number of five-branes discussed in section \ref{subsect:branecon}.
   In fact, on the supergravity side one computes the total number of D3-branes ending on any particular  five-brane {\emph {stack}}, so   the
   linking numbers of individual five-branes is actually defined by the ratios (\ref{linking}).
   This leads to the following definition of the partitions on the supergravity side:
   \begin{align}
\rho &= (\underbrace{l^{(1)},...,l^{(1)}}_{N^{(1)}_{D5}},...,\underbrace{l^{(a)},...,l^{(a)}}_{N^{(a)}_{D5}},...,\underbrace{l^{(q)},...,l^{(q)}}_{N^{(q)}_{D5}}) \ , \cr
\hat \rho &= (\underbrace{\hat l^{(1)},...,\hat l^{(1)}}_{\hat N^{(1)}_{NS5}},...,\underbrace{\hat l^{(b)},...,\hat l^{(b)}}_{\hat N^{(b)}_{NS5}},...,\underbrace{\hat l^{(\hat q)},...,\hat l^{(\hat q)}}_{\hat N^{(\hat q)}_{NS5}})  \ .
\label{branepartitions}
\end{align}
We note that the ordering chosen in (\ref{hmany}) for the location of the five-brane stacks $\delta_a$ and $\hat \delta_b$,
together with the expressions for the charges (\ref{gencharge}) and  the fact that
arctangent is a monotonic function,     implies that   $l^{(a)}$ and $\hat l^{(b)}$ have a canonical non-decreasing ordering
\beq
l^{(1)}\geq l^{(2)}\ldots>0 \ , \quad\qquad  \hat l^{(1)}\geq \hat l^{(2)}\ldots>0\ .
\eeq
From the  charge-conservation condition  (\ref{chargeconserv}) we furthermore  find:
 \bea
 \sum_{a=1}^q N^{(a)}_{D5} \, l^{(a)} \  = \ \sum_{b=1}^{\hat q} \hat N^{(b)}_{D3}\  \hat l^{(b)} \  =\ N\ ,
 \eea
 where we defined
 \bea
N \equiv  \sum_{a=1}^q   \sum_{b=1}^{\hat q}  N^{(a)}_{D5}\, \hat N^{(b)}_{NS5} \,  \frac{2}{\pi} \arctan (e^{\hat \delta_b - \delta_a})
\eea
Comparing the above expressions
with the parametrizations  (\ref{partitions}) and (\ref{partitionsb}) of the quiver data in section \ref{sec:quiver},  establishes
the basic gauge/gravity duality dictionary.
 \smallskip

An important remark is in order here:  the definition (\ref{linking}) of the linking numbers in the supergravity solution does not of course
require that these numbers be integers. This will only be the case if the number of D3-branes ending on the $a$-th D5-brane stack,
or on the $b$-th NS5-brane stack, is exactly divisible by the corresponding number of five-branes, respectively
$N^{(a)}_{D5}$ or $\hat N^{(b)}_{NS5}$.  The quantization of these latter numbers, or of the total numbers of D3-branes in a given stack,
are of course also not visible in supergravity.  They can be however deduced from a semi-classical Dirac-type argument in
the appropriate five-brane throat. The argument for quantization of the linking numbers  would have to be more subtle:
it would require splitting all asymptotic regions into individual five-brane throats.
\smallskip

Assuming the linking numbers to be integer, one notes
 that $N^{(a)}_{D5}$ is exactly the number of times the factor $l^{(a)}$ appears in \rf{branepartitions} while
 $\hat N^{(b)}_{NS5}$ is the number of times the factor  $\hat l^{(b)}$ appears in \rf{branepartitions}.
 Given our identification of the partitions of the supergravity solution  with
those of the dual superconformal  field theory, we arrive at the following identifications:
\begin{align}
\label{Mmap}
M_{l^{(a)}} = N^{(a)}_{D5} \qquad {\rm and} \qquad \hat M_{\hat l^{(b)}} = \hat N^{(b)}_{NS5} \ ,
\end{align}
where the $M_{l}$ and $\hat M_{\hat l }$ which do not explicitly appear in the above expressions
 are set to zero by default.
This  entry in the dictionary  identifies the numbers $M_{l^{(a)}}$ and $\hat M_{\hat l^{(b)}}$
of fundamental hypermultiplets coupled to each gauge group factor in  $T^{\rho}_{\hat \rho}(SU(N))$  and $T_{\rho}^{\hat \rho}(SU(N))$  respectively,
 with the number of D5-branes and NS5-branes in each five-brane stack  characterizing the corresponding type-IIB supergravity solution.


  \subsection{Bulk Realization of Fixed Point Symmetries}

Having completed the identification of parameters of the three dimensional field theories in our  supergravity solutions, the next step is to demonstrate that these latter precisely capture  the global symmetries of the conformal field theories labeled by $(\rho,\hat \rho)$.
As explained earlier, the superconformal symmetry $OSp(4 \vert 4)$ is manifest in the supergravity solution;
 the bosonic symmetries are  realized as isometries of the ${\rm AdS}_4\times {\rm S}^2\times  {\rm S}^2$ fibers.
  In fact the supergravity equations which determine the solutions
  were constructed by demanding that the type-IIB supergravity Killing
  spinor equations are satisfied for Killing spinors generating an $OSp(4 \vert 4)$ symmetry  \cite{Gomis:2006cu,DEG1,DEG2}.

The remaining task is to exhibit the rich global symmetry
\beq
H_{\rho}\times  H_{\hat \rho} =   \prod_i U(M_i)\times \prod_j U(\hat M_j)
 \eeq
of the superconformal theory in the corresponding supergravity solution. As has been explained in section \ref{sec:quiver},
this symmetry can be easily read  off from the manifest   flavour symmetry of the ultraviolet $T^{\rho}_{\hat \rho}(SU(N))$  and $T_{\rho}^{\hat \rho}(SU(N))$ quiver gauge theories which flow to this conformal field theory in the infrared.
The question is therefore, how can the $H_\rho\times H_{\hat\rho}$ global symmetry be realized
 in the corresponding  supergravity solution?

To answer this question,  recall that in holographic correspondences  conserved currents associated with global symmetries of the
boundary theory are associated to bulk gauge fields, and therefore to bulk gauge symmetries.
As we have rather explicitly demonstrated in section \ref{sec:soln}, our solutions  behave near the location of the singularities of the strip  as five-branes.
More precisely, the behaviour of the fields near a singularity in the upper/lower boundary of the strip is that due to a stack of D5/NS5--branes   with an $AdS_4\times S^2$ worldvolume.
The supergravity solution by itself is incomplete near these singularities. However, in string theory
the presence of
five-brane sources of precisely the required type, implies that near these sources we should place explicit five-branes in the geometry.  By usual string theory arguments involving the quantization of open strings ending on branes, new degrees of freedom are localized on these five-branes, and our supergravity solution must be enriched by taking them into account.
\smallskip

Among the  degrees of freedom introduced by a    stack of  $n$ coincident five-branes,
are $U(n)$ gauge fields supported on $AdS_4\times S^2$.
Therefore, taking into account that our supergravity solutions have $q$ stacks of D5-branes with $N^{(a)}_{D5}$ branes in each stack and
$\hat q$ stacks of D5-branes with $\hat N^{(b)}_{NS5}$ branes in each stack (see \ref{branepartitions}),
we find the following gauge symmetry
\bea
H_{\rho}\times  H_{\hat \rho}
 \ =\
 \prod_{a=1}^{q} U(N^{(a)}_{D5})  \times  \prod_{b =1}^{\hat q} U(\hat N^{(b)}_{NS5})\ .
\eea
The identification  \rf{Mmap} between the numbers of five-branes in every stack and the numbers  of fundamental hypermultiplets
in the ultraviolet $T^{\rho}_{\hat \rho}(SU(N))$  and $T_{\rho}^{\hat \rho}(SU(N))$ quiver gauge theories,
shows that the global symmetry of the field theory is precisely the gauge symmetry in the bulk solution.
The proposed holographic correspondence thus passes successfully this test.

 \subsection{Matching Constraints}
 As discussed in section \ref{sec:quiver}, in order for the $T^{\rho}_{\hat \rho}(SU(N))$
  theories to flow to a non-trivial infrared fixed point, the partitions $\rho$ and $\hat \rho$ must satisfy the condition
 $\rho^T > \hat \rho$.
  When  the bound is saturated, the theory becomes reducible.
    We shall now show that the supergravity solutions generally obey the constraint $\rho^T > \hat \rho$,
    except at certain degeneration limits where the bound is saturated.

To simplify the formulae leading to a proof of  this constraint on the gravity side,
 we first introduce the reduced notation
\begin{align}
N_a = N^{(a)}_{D5} \qquad ;  \qquad \hat N_b = \hat N^{(b)}_{NS5} \ .
\end{align}
Making use of the explicit expressions for the charges (\ref{genchargequant}), we  can  express
the linking numbers  $l^{(a)}$ and $l^{(b)}$ as
follows:
\begin{align}
\label{simpnotation}
l^{(a)} = \sum_{b = 1}^{\hat{q}} \hat N_b\,  f( \hat{\delta}_{b}-\delta_{a}) \qquad ;  \qquad
\hat l^{(b)} = \sum_{a = 1}^{q} N_a\,  f( \hat{\delta}_{b}-\delta_{a}) \ ,
\end{align}
where we also introduced the function
\beq
f(x) \equiv  \frac{2}{\pi} \arctan(e^{x}) \ .
\nonumber
\eeq

Recall that the
  partitions $\rho$ and $\hat\rho$ were defined in supergravity as
\begin{align}
\rho &= (\underbrace{l^{(1)},...,l^{(1)}}_{N_1},...,\underbrace{l^{(a)},...,l^{(a)}}_{N_a},...,\underbrace{l^{(q)},...,l^{(q)}}_{N_q}) \ ,  \cr
\hat \rho &= (\underbrace{\hat l^{(1)},...,\hat l^{(1)}}_{\hat N_1},...,\underbrace{\hat l^{(b)},...,
\hat l^{(b)}}_{\hat N_b},...,\underbrace{\hat l^{(\hat q)},...,\hat l^{(\hat q)}}_{\hat N_{\hat q}})\,.
\end{align}
The partition $\rho^T$  is then easily expressed as follows:
\begin{align}
\rho^T = (\underbrace{\sum_{a=1}^q N_a,...,\sum_{a=1}^q N_a}_{l^{(q)}},...,\underbrace{\sum_{a=1}^A N_a,...,\sum_{a=1}^A N_a}_{l^{(A)} - l^{(A+1)}},...,\underbrace{N_1,...,N_1}_{l^{(1)} - l^{(2)}}) \ ,
\label{trasposat}
\end{align}
where in   the $i$-th ``block" the sum ranges from $a=1$   to $a=A \equiv q - i+ 1$.

Our goal is to prove the set of  inequalities  $\rho^T > \hat \rho$ using
the explicit formulae (\ref{simpnotation}).
The meaning of $\rho^T > \hat \rho$ was defined previously  in (\ref{ineq}), and we repeat it here for the reader's convenience:
\bea
\label{ineqrep}
\sum_{s=1}^r m_s \ > \  \sum_{s=1}^r \hat l_s\qquad \forall r = 1,\ldots ,  l_1\, ,
 \eea
where the $m_s$ are the lengths of the rows of the Young tableau  $\rho^T$. As already noted in section 2,
these conditions imply in particular that $l_1 < \hat k$.
\smallskip

Using the formula (\ref{simpnotation}), the condition $l_1 < \hat k$ becomes
\beq
 \sum_{b=1}^{\hat q} \hat N_b f(\hat \delta_b - \delta_1) < \sum_{b=1}^{\hat q} \hat N_b \,.
 \eeq
Since $f(x) < 1$ for any  finite $x$, this
 inequality is manifestly valid.
Next, we turn to the remaining inequalities \rf{ineqrep}.  As a start let us  show  that it is sufficient to
prove  the inequalities in \rf{ineqrep}   for
\begin{align}
\label{defr}
r = \sum_{b=1}^{J} \hat N_b \qquad {\rm where} \qquad J = 1, 2 ,..., \hat q -1 \ .
\end{align}
 To see why, assume that  $r$ is in the   range $\sum_{b=1}^{J-1}\hat N_b < r \leq \sum_{b=1}^{J}\hat N_b$,
 for some $J = 1, 2 ,..., \hat q -1$.  Then
 if    \rf{ineqrep} is satisfied for all $r'<r$ but not  for $r$, it  will not be satisfied for
  $r'' = \sum_{b=1}^{J} \hat N_b$ either.  This is because
   $\hat l_s$  is constant for $s$ in the range $\sum_{b=1}^{J-1}\hat N_b < s \leq \sum_{b=1}^{J}\hat N_b$, while the
 integer $m_s$, which belongs to a
 non-decreasing sequence of integers,
 does not increase as $s$ ranges over the values $\sum_{b=1}^{J-1}\hat N_b < s \leq \sum_{b=1}^{J}\hat N_b$.
 Conversely, if the constraint is satisfied for $r''$ then
  it will be satisfied also  for $r$.
   We remark here that the limit of decoupled quivers, corresponding to disjoint brane configurations, is reached when the inequality is saturated for some value of $r$, with the saturation preserved for $r' > r$.  Following the logic
   of the previous argument, such an $r$ must be of the form $r = \sum_{b=1}^{J}\hat N_b$.
\smallskip

Let us now take a fixed $J$ with $1 \leq J \leq \hat q -1$.
 By summing over the number of rows in $\rho^T$, we can always find an integer $I$ such that
\begin{align}
l^{(I)} > r \geq l^{(I+1)}\ ,
\end{align}
where we take $l^{(0)} = +\infty$ and $l^{(q+1)} = 0$.
 We may then write the sum over $m_s$ as
\begin{align}
\sum_{s=1}^r m_s &= \sum_{A=I+1}^q \sum_{a=1}^{A} N_a \left(l^{(A)}-l^{(A+1)}\right) + \left(r - l^{(I+1)}\right) \sum_{a=1}^{I} N_a \cr
&= \sum_{a=I+1}^q l^{(a)} N_a + \left(\sum_{b=1}^{J} \hat N_b\right) \left(\sum_{a=1}^{I} N_a\right) \ ,
\end{align}
where we have used (\ref{defr}) to replace $r$.
The inequality (\ref{ineqrep}) then becomes
\beq
\sum_{b=1}^{J}\hat l^{(b)} \hat N_b < \sum_{a=I+1}^{q} l^{(a)} N_a
+ \left(\sum_{a=1}^{I} N_a \right)\left(\sum_{b=1}^{J} \hat N_b\right) \ .
\nonumber
\eeq
This is the form of the inequality that we will now prove using the supergravity calculation of the charges.
\smallskip

Making use of the expressions (\ref{simpnotation}) for the linking numbers, we can rewrite the above inequality as follows:
\beq
\sum_{a = 1}^{q} \sum_{b=1}^{J} N_a \hat N_b\,  f( \hat{\delta}_{b}-\delta_{a})
< \sum_{a=I+1}^{q} \sum_{b = 1}^{\hat{q}} N_a \hat N_b\,  f ( \hat{\delta}_{b}-\delta_{a} )  + \sum_{a=1}^{I}\sum_{b=1}^{J} N_a \hat N_b \ .
\nonumber
\eeq
Splitting the sums, simplifying and rearranging gives :
\beq
\sum_{a = 1}^{I} \sum_{b=1}^{J} N_a \hat N_b \, f( \hat{\delta}_{b}-\delta_{a})
\,  <\,  \sum_{a=1}^{I}\sum_{b=1}^{J} N_a \hat N_b + \sum_{a=I+1}^{q} \sum_{b = J+1}^{\hat{q}} N_a \hat N_b \, f( \hat{\delta}_{b}-\delta_{a}) \ .
\nonumber
\eeq
For finite values of $\delta_a$ and $\hat \delta_b$, this inequality is manifestly true because
$ 0 < f(x) < 1 $ for all finite $x$.
We notice that this inequality is saturated in two different limits: \\
 \indent (i) when $\delta_a \rightarrow +\infty$ for $a=I+1, I+2, ..., q$ and   $\hat \delta_b \rightarrow +\infty$ for $b=1, 2, ..., J$,
 or\\
  \indent (ii) when $\delta_a \rightarrow -\infty$ for $a=1, 2, ..., I$ and $\hat \delta_b \rightarrow -\infty$ for $j=J+1, J+2, ..., \hat q$.\\
  In the supergravity solution, these two  limits are related by a
   singular coordinate transformation corresponding to a large (infinite) translation of the strip.

 \subsection{Degeneration limits as wormholes}

 The limits in which  one or more of  the inequalities  contained in the statement $\rho^T > \hat\rho$
  become equalities, are of special significance.  As we have just seen these limits correspond, on the supergravity
  side, to  detaching  a subset of  five-brane singularities  and moving them off to infinity on the strip.
  On the field theory side, on the other hand,  the quiver gauge theory breaks up into two (or more) pieces,
  which are  connected
  by a  ``weak node", {\rm i.e.} a node of the quiver diagram  for which the gauge group has  rank   much much  smaller
  than the ranks of all  other gauge groups.  We will now make this statement  more explicit.

Consider the limit (i) in which
$\delta_a \rightarrow +\infty$ for $a=I+1, I+2, ..., q$ and   $\hat \delta_b \rightarrow +\infty$ for $b=1, 2, ..., J$
(the limit (ii) is as we have just argued equivalent). In this limit the charges (\ref{genchargequant})  for the five-brane stacks
at finite $z$ reduce to:
\begin{align}
N^{(a)}_{D3} &=- N_{D5}^{(a)} \sum_{b=1}^{J} \hat N^{(b)}_{NS5} - N^{(a)}_{D5}
\sum_{b=J+1}^{\hat q} \hat N^{(b)}_{NS5} \frac{2}{\pi} \arctan (e^{\hat \delta_b - \delta_a})  \ , \qquad a= 1, ..., I \cr
\hat N^{(b)}_{D3} &= \hat N_{NS5}^{(b)} \sum_{a=1}^{I} N^{(a)}_{D5} \frac{2}{\pi} \arctan (e^{\hat \delta_b - \delta_a})
 \ , \qquad \qquad   \qquad b= J+1,...,\hat q \ .
\end{align}
The extra contribution in $N^{(a)}_{D3}$ coming from the branes located at $\infty$ is actually irrelevant,
as it can be removed by an appropriate gauge transformation of $B_2$.
 This corresponds to choosing the gauge so that $B_2 =0$  on the segment $(\delta_I,\delta_{I+1})$.
In this way,  a  solution   with $I$   D5-branes stacks and $(\hat q- J)$  NS5-brane stacks is detached from the rest
of the geometry.
\smallskip

\begin{figure}
\center
\vskip -1.2cm
\includegraphics[width=12cm]{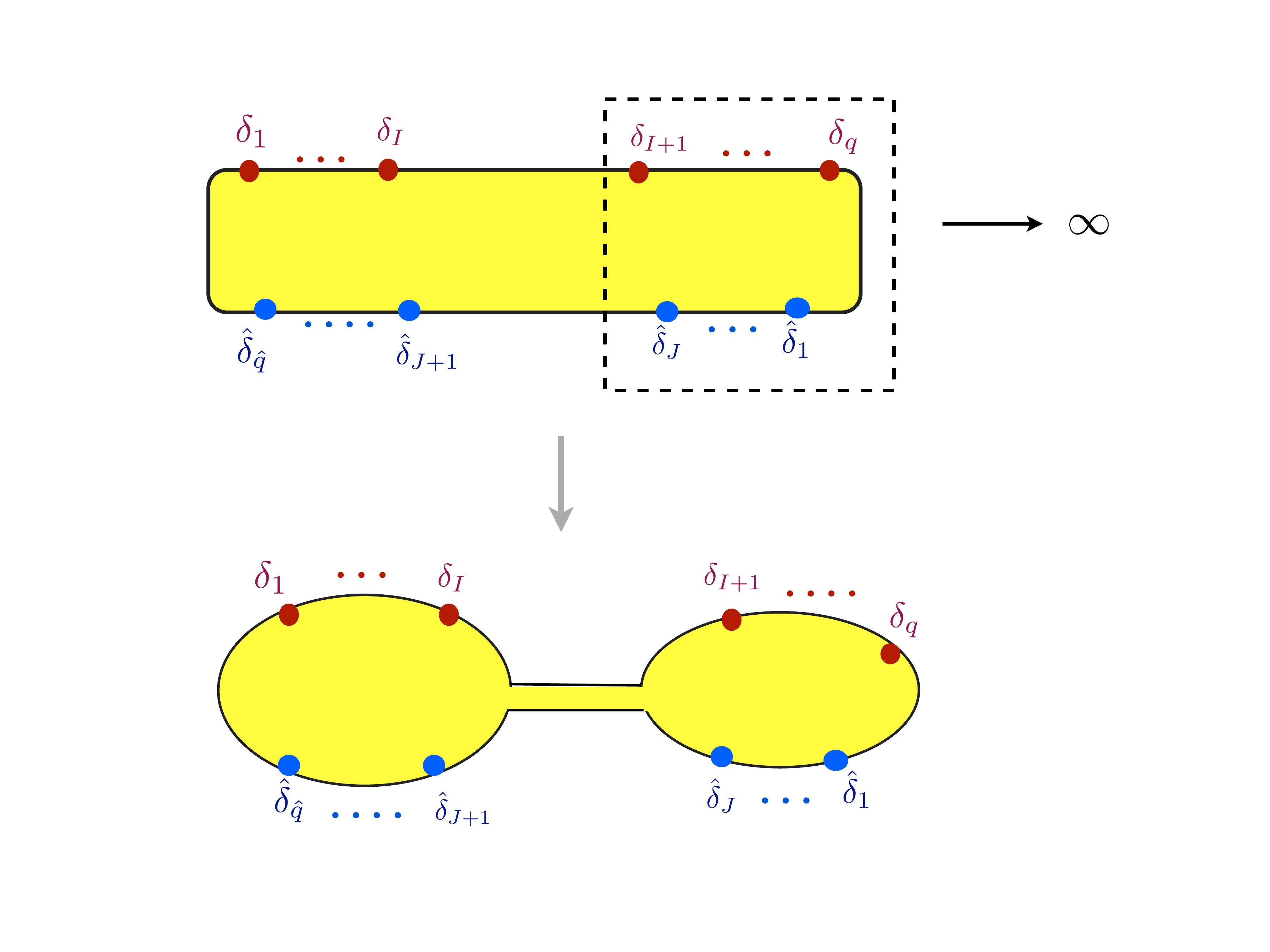}
\vskip -1.2cm
\caption{\footnotesize
Schematic drawing of the factorization limit of five-brane singularities discussed in the text. The picture
is meant to show the actual size of the strip geometry.  The background consists of two,    $AdS_4\ltimes K$
and   $AdS_4\ltimes K^\prime$,   solutions
coupled through a narrow $AdS_5\times S^5$ bridge.  The curvature of the narrow bridge is larger
than the curvature in the rest of the  geometry, but can be small enough so as  to ignore quantum gravity corrections.
The configuration resembles therefore a wormhole.
  }
\label{factorize}
\end{figure}

 More generally, if we also keep track of the five-branes moving off to infinity, we find a supergravity solution
 which consists of two geometries of type $AdS_4\ltimes K$ and
 $AdS_4\ltimes K^\prime$, connected by a narrow bridge, as illustrated in  figure \ref{factorize}.
 The space   $AdS_4\ltimes K$ corresponds to
  keeping  only the stacks $a = 1 , 2 , ... , I$ , $ b = J+1, J+2, ..., \hat q$,
  while  the space   $AdS_4\ltimes K^\prime$  is the solution obtained if we only keep
  the five-brane stacks $a = I+1 , I+2 , ... , q$ ,  and $ b = 1, 2, ..., J$.  Saturating the relation $\rho^T \geq \hat \rho$
  corresponds to eliminating all  D3-branes in the intermediate region.  It can be checked indeed that,  in the
  limit,  the D3-brane charge is separately conserved in the two regions.

We can check that the partitions corresponding to these two solutions are exactly the ones obtained by the splitting of $(\rho,\hat \rho)$ into two subpartitions by the saturation of the condition (\ref{ineq}) for $r=\sum_{b=1}^{J}\hat N_b$. These partitions are explicitly :
\bea
\rho_{L} &=& \Big( \underbrace{l_1 -\sum_{b=1}^{J} \hat N_b,...,l_1 - \sum_{b=1}^{J} \hat N_b}_{N_1} , ... , \underbrace{l_I - \sum_{b=1}^{J} \hat N_b,...,l_I - \sum_{b=1}^{J} \hat N_b}_{N_{I}} \Big)
\nonumber\\
\hat{\rho}_{L} &=& \Big(\underbrace{\hat{l}_{J+1},...,\hat{l}_{J+1}}_{\hat{N}_{J+1}}, ... ,\underbrace{\hat{l}_{\hat{q}},...,\hat{l}_{\hat{q}}}_{\hat{N}_{\hat{q}}}  \Big)
\eea
and
\bea
\rho_{R} &=& \Big( \underbrace{l_{I+1},...,l_{I+1}}_{N_{I+1}} , ... , \underbrace{l_{q},...,l_{q}}_{N_{q}} \Big)
\nonumber\\
\hat{\rho}_{R} &=& \Big(\underbrace{\hat{l}_{1} - \sum_{a=1}^{I} N_{a},...,\hat{l}_{1} - \sum_{a=1}^{I} N_{a}}_{\hat{N}_{1}}, ... ,\underbrace{\hat{l}_{J} - \sum_{a=1}^{I} N_{a},...,\hat{l}_{J} - \sum_{a=1}^{I} N_{a}}_{\hat{N}_{J}}  \Big)
\eea
where the indices L, R refer to the left and right parts of the split quiver.
 The   linking numbers have been here gauge transformed  so as   to make them agree, for each sub-quiver separately,  with our
 earlier conventions.
  So the splitting of the  quiver  corresponds precisely to the factorization of the bulk geometry, confirming
  once again the holographic duality map.

As we have seen in section \ref{sec:soln},
  the limit of capping off asymptotic $AdS_5\times S^5$ regions  is smooth. We hope to return to the
  physics of this limit elsewhere.


   \section{Discussion}
  \label{sec:discuss}

In this paper we have constructed the type-IIB  supergravity solutions which are holographically dual to  a rich family of  three dimensional $\cN=4$ superconformal field theories.
These  theories arise as infrared fixed points of the   $T^\rho_{\hat \rho}(SU(N))$  and $T_\rho^{\hat \rho}(SU(N))$ quiver gauge theories whenever $\hat\rho^T>\rho$. This non-trivial constraint, together with the  $H_\rho\times H_{\hat\rho}$ global symmetries of the associated superconformal field theory, have been precisely realized in our supergravity solutions.

Our explicit type-IIB supergravity solutions provide a novel arena in which to study this rich family of superconformal field theories.\footnote{For recent work on infrared fixed points in $N = 2$ Chern-Simons matter theories see \cite{Bianchi:2009ja,Bianchi:2009rf}.}
Even though the dilaton or curvature gets large near the location of five-brane singularities, our solution can be nicely interpreted in string theory by replacing the five-brane
singularities by explicit five-branes, which give rise to important new light degrees of freedom localized in the geometry. Information regarding the spectrum of local and non-local operators in these conformal field theories can be obtained by studying the supergravity fluctuation spectrum around our $AdS_4\ltimes K$ solutions as well as by considering strings and branes ending on the boundary of our $AdS_4\ltimes K$ backgrounds along submanifolds of varying dimensionality.

Another very interesting direction
 is to use our supergravity  solutions to determine the partition function $Z_{S^3}$ of the boundary field theory on $S^3$, obtained by evaluating the type-IIB string action on the $AdS_4\ltimes K$ solutions. Recently, it has been noted that the associated renormalized ``free energy"  \cite{Casini:2011kv,Jafferis:2011zi} (see also \cite{Amariti:2011da})
\beq
F=-\log |Z_{S^3}|
\eeq
enjoys interesting monotonicity properties under renormalization group evolution.\footnote{This observable of three dimensional field theories is a close cousin to the conformal anomaly coefficient $a$ of four dimensional field theories, which is also conjectured to decrease along renormalization group trajectories and to be stationary at fixed points.} The partition function  $Z_{S^3}$  of the infrared superconformal field theory associated to $T^\rho_{\hat \rho}(SU(N))$ deformed by FI and mass parameters has recently  been calculated \cite{Nishioka:2011dq} (see also \cite{Benvenuti:2011ga}) using the localization formulae in \cite{Kapustin:2009kz}, and shown to reproduce the
partition function of the mirror $T_\rho^{\hat \rho}(SU(N))$  theory upon exchanging the role of FI and mass parameters. By suitably taking the deformation parameters to their ``superconformal" value, the formula for the partition function at the superconformal fixed point can be obtained, and compared with the one calculated from our supergravity solutions.\footnote{Analogous comparisons have been successfully performed for a different family of three dimensional superconformal field theories which have M-theory gravitational dual descriptions of the type $AdS_4\times X_7$ (see e.g \cite{Drukker:2010nc}\cite{Martelli:2011qj}\cite{Cheon:2011vi}\cite{Gulotta:2011si}).}

Also, as we have seen in this paper, the above type-IIB geometries have  interesting factorization limits, as well as limits in which
 asymptotic $AdS_5\times S^5$ regions become  very highly curved.
The former can be thought of as  wormhole-like  solutions  which describe two different $AdS_4\ltimes K$ regions,
  connected by  an $AdS_5\times S^5$ throat, while in the latter limit  a large $AdS_4\ltimes K$ region  is extended to infinity along one or more
 very thin $AdS_5\times S^5$ fixtures or throats.
We plan to return to the physics of these solutions,  and whether they give a consistent string theory realization of massive gravity or multi-gravity.

Finally, we would like to point out that the  solutions of type-IIB string theory constructed  in this paper have no moduli!
  That is, the quantization
condition of the various fluxes, and   the presence of both NS5 and D5-branes in the  geometry,  fix  all moduli, including the dilaton.
It is interesting that rather simple and explicit isolated vacua of string theory can be explicitly constructed. It would be desirable to
determine whether flux quantization in the presence of both NS5 and D5-branes can be used to construct
phenomenologically more realistic vacua of string theory.

       \vskip 0.5cm

{\bf Aknowledgements}:
{We thank  D. Gaiotto and Y. Tachikawa for   discussions.
C.B. thanks the Alexander von Humboldt foundation  and the Ludwig Maximilian
Universit\"at in M\"unich for hospitality in the final stages  of this work.
J.E. is supported by the FWO - Vlaanderen, Project No. G.0235.05, and by
 the ``Federal Office for Scientific, Technical and Cultural Affairs through the Inter-University Attraction Poles
 Programme,"  Belgian Science Policy P6/11-P.   J.G. thanks the LPTENS, the LPTHE in Jussieu and the
FRIF  (``Federation de Recherche sur les Interactions Fondamentales") for their hospitality during
 this work.  J.G. further thanks the University of Barcelona for hospitality during the completion of this work.
 Research at the Perimeter Institute is supported in part by the Government of Canada through NSERC and by the Province of Ontario through MRI.
J.G. also acknowledges further support from an NSERC Discovery Grant and from an ERA grant by the Province of Ontario. }


\vfil\eject



\begin{thebibliography}{99}

\bibitem{Maldacena:1997re}
  J.~M.~Maldacena,
  ``The Large N limit of superconformal field theories and supergravity,''
  Adv.\ Theor.\ Math.\ Phys.\  {\bf 2 } (1998)  231-252.
  [hep-th/9711200].

\bibitem{Gaiotto:2008ak}
  D.~Gaiotto, E.~Witten,
  ``S-Duality of Boundary Conditions In N=4 Super Yang-Mills Theory,''
   [arXiv:0807.3720 [hep-th]].


\bibitem{Nishioka:2011dq}
  T.~Nishioka, Y.~Tachikawa and M.~Yamazaki,
  ``3d Partition Function as Overlap of Wavefunctions,''
  arXiv:1105.4390 [hep-th].

\bibitem{DEG1}
  E.~D'Hoker, J.~Estes and M.~Gutperle,
  ``Exact half-BPS Type IIB interface solutions I: Local solution and supersymmetric Janus,''
  JHEP {\bf 0706} (2007) 021
  [arXiv:0705.0022 [hep-th]].

\bibitem{DEG2}
  E.~D'Hoker, J.~Estes and M.~Gutperle,
  ``Exact half-BPS type IIB interface solutions. II: Flux solutions and multi-janus,''
  JHEP {\bf 0706} (2007) 022
  [arXiv:0705.0024 [hep-th]].

\bibitem{Gomis:2006cu}
  J.~Gomis and C.~Romelsberger,
  ``Bubbling defect CFT's,''
  JHEP {\bf 0608} (2006) 050
  [arXiv:hep-th/0604155].

\bibitem{Lunin:2006xr}
  O.~Lunin,
  ``On gravitational description of Wilson lines,''
  JHEP {\bf 0606}, 026 (2006).
  [hep-th/0604133].


\bibitem{Gaiotto:2008sa}
  D.~Gaiotto, E.~Witten,
  ``Supersymmetric Boundary Conditions in N=4 Super Yang-Mills Theory,''
  [arXiv:0804.2902 [hep-th]].

\bibitem{Bachas:2011xa}
  C.~Bachas, J.~Estes,
  ``Spin-2 spectrum of defect theories,''
  JHEP {\bf 1106 } (2011)  005.
  [arXiv:1103.2800 [hep-th]].


\bibitem{Aharony:2003qf}
  O.~Aharony, O.~DeWolfe, D.~Z.~Freedman,  A.~Karch,
  ``Defect conformal field theory and locally localized gravity,''
  JHEP {\bf 0307}, 030 (2003).
  [hep-th/0303249].



\bibitem{Kiritsis:2006hy}
  E.~Kiritsis,
  ``Product CFTs, gravitational cloning, massive gravitons and the space of
  gravitational duals,''
  JHEP {\bf 0611} (2006) 049
  [arXiv:hep-th/0608088].

\bibitem{Aharony:2006hz}
  O.~Aharony, A.~B.~Clark and A.~Karch,
  ``The CFT/AdS correspondence, massive gravitons and a connectivity index
  conjecture,''
  Phys.\ Rev.\  D {\bf 74} (2006) 086006
  [arXiv:hep-th/0608089].

\bibitem{Aharony:2011yc}
  O.~Aharony, L.~Berdichevsky, M.~Berkooz, I.~Shamir,
  ``Near-horizon solutions for D3-branes ending on 5-branes,''
  [arXiv:1106.1870 [hep-th]].



\bibitem{Intriligator:1996ex}
  K.~A.~Intriligator, N.~Seiberg,
  ``Mirror symmetry in three-dimensional gauge theories,''
  Phys.\ Lett.\  {\bf B387}, 513-519 (1996).
  [hep-th/9607207].

\bibitem{Hanany:1996ie}
  A.~Hanany, E.~Witten,
  ``Type IIB superstrings, BPS monopoles, and three-dimensional gauge dynamics,''
  Nucl.\ Phys.\  {\bf B492}, 152-190 (1997).
  [hep-th/9611230].





\bibitem{Bachas:1997ui}
  C.~P.~Bachas, M.~R.~Douglas, M.~B.~Green,
 ``Anomalous creation of branes,''
  JHEP {\bf 9707}, 002 (1997).
  [hep-th/9705074].

\bibitem{Bachas:1997kn}
  C.~P.~Bachas, M.~B.~Green and A.~Schwimmer,
  ``(8,0) quantum mechanics and symmetry enhancement in type I' superstrings,''
  JHEP {\bf 9801} (1998) 006
  [arXiv:hep-th/9712086].


\bibitem{Marolf:2000cb}
  D.~Marolf,
  ``Chern-Simons terms and the three notions of charge,''
   [hep-th/0006117].


\bibitem{BDS}
  C.~Bachas, M.~R.~Douglas, C.~Schweigert,
  ``Flux stabilization of D-branes,''
  JHEP {\bf 0005}, 048 (2000).
  [hep-th/0003037].

\bibitem{Bianchi:2009ja}
 M.~S.~Bianchi, S.~Penati and M.~Siani,
  ``Infrared stability of ABJ-like theories,''
  JHEP {\bf 1001} (2010) 080
  [arXiv:0910.5200 [hep-th]].

\bibitem{Bianchi:2009rf}
  M.~S.~Bianchi, S.~Penati and M.~Siani,
  ``Infrared Stability of N = 2 Chern-Simons Matter Theories,''
  JHEP {\bf 1005} (2010) 106
  [arXiv:0912.4282 [hep-th]].



\bibitem{Casini:2011kv}
  H.~Casini, M.~Huerta, R.~C.~Myers,
``Towards a derivation of holographic entanglement entropy,''
  JHEP {\bf 1105}, 036 (2011).
  [arXiv:1102.0440 [hep-th]].

\bibitem{Jafferis:2011zi}
  D.~L.~Jafferis, I.~R.~Klebanov, S.~S.~Pufu, B.~R.~Safdi,
  ``Towards the F-Theorem: N=2 Field Theories on the Three-Sphere,''
  [arXiv:1103.1181 [hep-th]].

\bibitem{Amariti:2011da}
 A.~Amariti and M.~Siani,
  ``Z-extremization and F-theorem in Chern-Simons matter theories,''
  arXiv:1105.0933 [hep-th]. 

\bibitem{Benvenuti:2011ga}
  S.~Benvenuti and S.~Pasquetti,
  ``3D-partition functions on the sphere: exact evaluation and mirror symmetry,''
  arXiv:1105.2551 [hep-th].


\bibitem{Kapustin:2009kz}
  A.~Kapustin, B.~Willett, I.~Yaakov,
  ``Exact Results for Wilson Loops in Superconformal Chern-Simons Theories with Matter,''
  JHEP {\bf 1003}, 089 (2010).
  [arXiv:0909.4559 [hep-th]].

\bibitem{Drukker:2010nc}
  N.~Drukker, M.~Marino, P.~Putrov,
  ``From weak to strong coupling in ABJM theory,''
  [arXiv:1007.3837 [hep-th]].

\bibitem{Martelli:2011qj}
  D.~Martelli, J.~Sparks,
   ``The large N limit of quiver matrix models and Sasaki-Einstein manifolds,''
   [arXiv:1102.5289 [hep-th]].

\bibitem{Cheon:2011vi}
  S.~Cheon, H.~Kim, N.~Kim,
  ``Calculating the partition function of N=2 Gauge theories on $S^3$ and AdS/CFT correspondence,''
  JHEP {\bf 1105}, 134 (2011).
  [arXiv:1102.5565 [hep-th]].

\bibitem{Gulotta:2011si}
  D.~R.~Gulotta, C.~P.~Herzog, S.~S.~Pufu,
  ``From Necklace Quivers to the F-theorem, Operator Counting, and T(U(N)),''
   [arXiv:1105.2817 [hep-th]].




\end{thebibliography}
\end{document}